\documentclass{aa}  

\usepackage{graphicx}
\usepackage{txfonts}
\usepackage{amssymb}
\usepackage{amsmath}
\usepackage{graphicx}
\usepackage[colorlinks=true, allcolors=blue]{hyperref}
\usepackage{booktabs}
\usepackage{amsfonts}     
\usepackage[T1]{fontenc} 
\usepackage[utf8]{inputenc} 
\usepackage{multicol}
\usepackage{xcolor}
\usepackage[switch]{lineno}
\usepackage{gensymb}
\usepackage{soul}
\usepackage{rotating}
\usepackage{comment}
\usepackage{multirow}
\usepackage{longtable}
\usepackage{lscape}
\usepackage{color}
\usepackage{placeins}

\newcommand{\eqb}{\begin{eqnarray}}
\newcommand{\eqe}{\end{eqnarray}}

\newcommand{\subsubsubsection}[1]{\paragraph{#1}\mbox{}\\}
\begin{document}

\title{CAZ catalog and optical light curves of 7918 blazar-selected active galactic nuclei}

\subtitle{}

\author{
Pouya M. Kouch \inst{\ref{UTU},\ref{FINCA},\ref{MRO}} \thanks{\href{mailto:pouya.kouch@utu.fi}{pouya.kouch@utu.fi}} \orcid{0000-0002-9328-2750},
Elina Lindfors \inst{\ref{UTU},\ref{FINCA}} \orcid{0000-0002-9155-6199}, 
Talvikki Hovatta \inst{\ref{FINCA},\ref{MRO},\ref{ELE}} \orcid{0000-0002-2024-8199}, 
Ioannis Liodakis \inst{\ref{IA_FORTH_Crete}} \orcid{0000-0001-9200-4006}, 
Karri I.I. Koljonen \inst{\ref{NTNU}} \orcid{0000-0002-9677-1533}, 
Alessandro Paggi \inst{\ref{IA_FORTH_Crete}} \orcid{0000-0002-5646-2410}, 
Kari Nilsson \inst{\ref{FINCA}} \orcid{0000-0002-1445-8683}, 
Jenni Jormanainen \inst{\ref{UTU},\ref{FINCA}} \orcid{0000-0003-4519-7751}, 
Vandad Fallah Ramazani \inst{\ref{FINCA}} \orcid{0000-0001-8991-7744}, 
Sofia Kankkunen \inst{\ref{MRO},\ref{ELE}} \orcid{0009-0008-0460-7826}, 
Folkert Wierda \inst{\ref{UTU},\ref{FINCA}} \orcid{0000-0002-8657-0076}, 
Sarah M. Wagner \inst{\ref{Wurzburg}} \orcid{0000-0002-8423-6947}, 
and
Matthew J. Graham \inst{\ref{CalTech}} \orcid{0000-0002-3168-0139} 
}

\institute{
Department of Physics and Astronomy, University of Turku, FI-20014 Turku, Finland \label{UTU}
\and
Finnish Centre for Astronomy with ESO (FINCA), Quantum, Vesilinnantie 5, University of Turku, FI-20014 Turku, Finland \label{FINCA}
\and
Aalto University Mets\"ahovi Radio Observatory, Mets\"ahovintie 114, FI-02540 Kylm\"al\"a, Finland \label{MRO}
\and
Aalto University Department of Electronics and Nanoengineering, PL~15500, FI-00076 Espoo, Finland\label{ELE}
\and
Institute of Astrophysics, Foundation for Research and Technology-Hellas, GR-71110 Heraklion, Greece \label{IA_FORTH_Crete}
\and
Institutt for Fysikk, Norwegian University of Science and Technology, H{\o}gskloreringen 5, Trondheim, 7491, Norway \label{NTNU}
\and
Institut f\"ur Theoretische Physik and Astrophysik, Universit\"at W\"urzburg, D-97074 W\"urzburg, Germany \label{Wurzburg}
\and
Division of Physics, Mathematics and Astronomy, California Institute of Technology, Pasadena, CA91125, USA \label{CalTech}
}

\date{Received October 07, 2025; accepted March 06, 2026}

 
\abstract{Active galactic nuclei (AGN) are some of the brightest and most variable objects in the Universe. Those with relativistic jets observed at small viewing angles are blazars. Due to Doppler boosting, blazars exhibit extreme stochastic variability. While the origin of this variability is thought to be changes in the accretion flow and jet dynamics, much about blazar variability remains unknown. In this paper we use several blazar-dominated AGN samples to form a catalog of 7918 blazars and candidates -- the largest to date. We also collected source types, redshifts, peak frequencies of the spectral energy distribution, radio variability Doppler factors, and X-ray flux densities for as many sources as possible. We used all-sky surveys (CRTS, ATLAS, and ZTF, abbreviated as ``CAZ'') to extract their optical multiband flux density on a nightly basis between 2007 and 2023, and we constructed as long and as high cadence light curves as possible for as many sources as attainable. We quantified the variability of the light curves and applied the Bayesian blocks algorithm to determine their flaring periods. The CAZ catalog and light curves as well as the corresponding Bayesian blocks and flaring periods are all provided in the accompanying electronic tables, with the goal of enabling analyses involving jetted AGN variability with unprecedented sample sizes. Overall, we find (1) optical flares generally have a faster rise than decay; (2) optical brightness and variability are strongly dependent on the synchrotron peak frequency; (3) flat spectrum radio quasars and BL Lac objects have comparable optical variability and flare characteristics at the same synchrotron peak frequency; and (4) optical flare times tend to decrease while amplitudes increase with an increasing radio variability Doppler factor.}

\keywords{catalogs - galaxies: active - galaxies: jets - galaxies: statistics - galaxies: quasars: general - galaxies: BL Lacertae objects: general}

\titlerunning{CAZ catalog and optical light curves of 7918 blazar-selected AGN}
\authorrunning{Kouch et al.}

\maketitle
%

\section{Introduction} \label{sec_intro}
Active galactic nuclei (AGN) exhibit significant non-stellar emission powered by supermassive black holes (SMBH) in the center of their host galaxy. Around a tenth of AGN possess highly collimated jets of relativistic plasma. Blazars constitute a small fraction of jetted AGN, whose jet points toward Earth at small viewing angles ($\lesssim$10$^\circ$). This leads to a relativistically boosted broadband nonthermal emission from radio to very-high-energy (0.1--100~TeV) $\gamma$-rays, making blazars some of the brightest extragalactic objects in the Universe. For recent reviews on blazars, see \cite{blandford2019}, \cite{Bottcher2019}, and \cite{hovatta_lindfors2019}. 

The energy output of blazars in terms of frequency is often studied via their spectral energy distribution (SED). All blazar SEDs show a double-hump structure, the first of which is due to synchrotron radiation of relativistic electrons (typically peaking between the radio and ultraviolet bands). The second hump typically peaks at the X-ray or higher energy bands, and its origin is not fully understood (generally dominated by leptonic inverse Compton scattering, but hadronic contributions cannot be ruled out; e.g., \citealt{Marshall2024_IXPE_LSPs, Agudo2025_IXPE_hadronic_disfavored, Liodakis2025_IC_vs_hadronic}). Blazars are often categorized based on their synchrotron peak frequency into low-, intermediate-, and high-synchrotron peaked sources, respectively LSP ($\nu_{\mathrm{sy}}<10^{14}$~Hz), ISP ($10^{14} \leq \nu_{\mathrm{sy}}<10^{15}$~Hz), and HSP ($\nu_{\mathrm{sy}} \geq 10^{15}$~Hz; e.g., \citealt{abdo_2010_synch_peak_classification}).

A more traditional classification of blazars is based on observational features. Flat spectrum radio quasars (FSRQs) are those whose optical spectra exhibit broad emission lines -- rest-frame equivalent width (EW) of $>$5~\AA  -- and are typically LSPs. On the other hand, BL Lac objects (BLLs) display mostly featureless spectra (EW~$<$~5~\AA), and their synchrotron emission can peak in a wide range of energy bands, making them LSPs, ISPs, HSPs, or even extreme HSPs (EHSPs; where $\nu_{\mathrm{sy}} \geq 10^{17}$~Hz; e.g., \citealt{costamente2001_extreme_hsp}). The FSRQs and BLLs are thought to constitute intrinsically different populations, at least partially (e.g., \citealt{Ghisellini2010_jet_power}). For example, FSRQs are known to have a broad-line region in the vicinity of their central engine where fast-moving gas clouds absorb and rescatter the light from the accretion disk. On the other hand, the central engine of BLLs is thought to be ``naked.'' As a result, seed photons for inverse Compton scattering are thought to come from the dusty torus, the broad-line region, or the accretion disk in FSRQs and from synchrotron photons in BLLs. Furthermore, FSRQs and BLLs show different SED behaviors as luminosity increases. In FSRQs, the SED peak frequencies remain constant, with the higher-energy component becoming brighter, whereas in BLLs both SED peaks shift to higher energies (e.g., \citealt{ghisellini2017_blz_sequence}).

The energy output of blazars, and more generally AGN, exhibits extreme stochastic variability over time (for a recent review, see, e.g., \citealt{Kankkunen2024_AGN_var_review_and_caveats}). As such, their light curves show periods of quiescent and enhanced activity (i.e., ``flaring''), during which the energy output can be tens of times greater than the quiescent period. Such flaring periods occur in all bands of the electromagnetic spectrum, although their typical characteristics vary greatly from band to band. For instance, in the radio band, they can be as long as years (e.g., \citealt{Hovatta2007_radio_longerm_var_analysis, Kankkunen2024_MRO_lognterm_var_analysis}) but as fast as minutes in higher energy bands, such as $\gamma$-rays (e.g., \citealt{Aharonian2007_VHE_minute_scale_var, Albert2007_Mrk501_minute_TeV_flare, Pandey2022_Fermi_minute_scale_var}). In the optical band, blazars show a wide range of timescales, from years to minutes (e.g., \citealt{Romero2002_opt_intranight_var, Ruan2012_typical_longtem_opt_var_timescale, Weaver2020_opt_fast_var_timescale, Jorstad2022_QPO_BLLac_Nature}). In general, multiwavelength blazar light curves can be used to investigate the overall jet activity over time (e.g., \citealt{Lahteenmaki2003_radio_gamma_correlation, leon_tavares2011_radio_gamma_correlation}), making them a powerful tool for answering some of the biggest open questions about blazars. For example, they are necessary for understanding the origin of the blazar high-energy emission (e.g., \citealt{Middei2023_IXPE_BLLac, Kouch2025_IXPE_0954}) and constraining the different acceleration mechanisms in the jet (e.g., \citealt{Liodakis2022_IXPE_nature, Kouch2024_PKS2155}); for more examples, see the end of the paper (Sect. \ref{sec_conclusion}). 

In this work, we compile a sample of 7918 AGN (mostly blazars; see Sect. \ref{sec_compiling_caz_sources}) and obtain long-term (as long as $\sim$15~yr) and high-cadence (as dense as nightly) flux density light curves in the optical band for as many of them as possible. The light curves are mainly extracted from three all-sky surveys (Sect. \ref{sec_lc_extraction}): the Catalina Real-time Transient Survey (CRTS), the Asteroid Terrestrial-impact Last Alert System (ATLAS), and the \textit{Zwicky} Transient Facility (ZTF) survey. Together, these surveys form the acronym ``CAZ'' (CRTS+ATLAS+ZTF), which we use to refer to our catalog of 7918 blazar-dominated AGN and their light curves. This catalog constitutes the most extensive and uniformly compiled sample of blazar optical light curves to date, making it ideal for use in population-based analyses. Nevertheless, we caution that some of the light curves suffer from various complications arising due to automatic forced-photometry (e.g., underestimated errors, presence of extreme outliers, and contamination by their host galaxy and bright nearby sources). In this work we also characterize the variability of the CAZ light curves and identify their flaring periods (Sect. \ref{sec_characterizing_variability}), whose results are discussed and interpreted in Sect. \ref{sec_results}. Our conclusions are summarized in Sect. \ref{sec_conclusion}. Our primary motivation to construct this catalog and its light curves has been to perform the largest-ever population-based spatio-temporal correlation between the blazar and high-energy neutrino populations (e.g., \citealt{hovatta2021, Kouch2024_CGRaBS_update, Kouch2025_WH2}), which we do in our companion paper, \cite{Kouch2025_companion_CAZ_v_IC}.

\section{Compiling the CAZ catalog} \label{sec_compiling_caz_sources}
We obtain long-term optical light curves for as many blazars and candidates as possible by compiling sources from blazar-dominated catalogs. We begin with the Very Long Baseline Interferometry (VLBI) based Radio Fundamental Catalog (RFC), which contains $\sim$22k precisely localized compact radio sources. We use the \texttt{rfc\_2023d}\footnote{\href{https://astrogeo.org/sol/rfc/rfc_2023d/}{https://astrogeo.org/sol/rfc/rfc\_2023d/}} version released in January 2024. Due to the efforts of the VCS (Very Long Baseline Array calibrator surveys; e.g., \citealt{kovalev2007_vcs5,gordon2016_vcsii}) and LBA (Australian Long Baseline Array; e.g., \citealt{petrov2019_lba}) observations, the subset of 3236 RFC sources fulfilling $S_\text{X-band}^{\text{VLBI}} \geq 150$~mJy can be considered statistically complete (i.e., this subset contains all compact sources detectable above 150~mJy in the X-band, 8--12~GHz). Since most sources in any flux-limited VLBI sample are expected to be blazars (e.g., \citealt{readhead1978_flux_limited_sample_mostly_blz}), we begin compiling the CAZ catalog using this subset of 3236 sources. We use this flux-limited subsample over the entire RFC catalog because: (1) it mostly contains confirmed blazars while the entire RFC catalog mostly contains sources of uncertain type, likely several thousand non-blazars; (2) using the entire RFC catalog would not benefit us when constructing the CAZ optical light curves since most of its sources would be too dim in the optical band; and (3) focusing on the flux-limited subsample enables population-based analyses on a statistically complete sample of radio-bright AGN.

Subsequently, we use the latest release of the Fourth Catalog of Active Galactic Nuclei present in the Large Array Telescope (4LAC), 4LAC-DR3,\footnote{\href{https://fermi.gsfc.nasa.gov/ssc/data/access/lat/4LACDR3/}{https://fermi.gsfc.nasa.gov/ssc/data/access/lat/4LACDR3/}} which was released in August 2022 and is based on 12 years of data from the \textit{Fermi} Large Array Telescope (LAT; \citealt{Atwood2009_FermiLAT}). We add all 3814 low- and high-latitude 4LAC sources into the CAZ source list, using their radio counterpart coordinates provided in the 4LAC catalog (\citealt{ajello2022_4lac_dr3}). Additionally, as seen below, we tabulate several other parameters (e.g., redshift, SED peaks) from the 4LAC catalog. We then use the third catalog of HSP blazars (3HSP\footnote{\href{https://www.ssdc.asi.it/3hsp/}{https://www.ssdc.asi.it/3hsp/}}), last updated in April 2020. We add all its 2013 sources into the CAZ source list, while adopting the counterpart coordinates. As discussed in \cite{chang2019_3hsp}, most of the counterparts in the 3HSP catalog are from the WISE\footnote{\href{https://wise2.ipac.caltech.edu/docs/release/allsky/}{https://wise2.ipac.caltech.edu/docs/release/allsky/}} all-sky survey (Wide-field Infrared Survey Explorer at IPAC); however, some only have 5BZC\footnote{\href{https://www.ssdc.asi.it/bzcat/}{https://www.ssdc.asi.it/bzcat/}} (the fifth Roma-BZCAT Multifrequency Catalogue of Blazars) or \textit{Fermi}-LAT (3LAC) counterparts. We tabulate the synchrotron peak frequency and redshift of the sources whenever provided by the 3HSP catalog. Next, we add all 1625 sources in the Candidate Gamma-Ray Blazar Survey (CGRaBS\footnote{\href{https://heasarc.gsfc.nasa.gov/W3Browse/radio-catalog/cgrabs.html}{https://heasarc.gsfc.nasa.gov/W3Browse/radio-catalog/cgrabs.html}}; \citealt{healey2008_cgrabs}) source catalog, which is complete down to 65~mJy flux density at 4.8~GHz and radio spectral index $\alpha > -0.5$ where $S \propto \nu^{\alpha}$ (\citealt{richards2011_ovro}). We adopt the redshift and source classification of this catalog. Finally, we add all 3561 sources of the aforementioned 5BZC catalog of blazars, while tabulating their redshift and 5BZC classification (\citealt{massaro2015_5bzc}).

In the next step, we cross-check the radio-counterpart RA and Dec. coordinates of the sources to the closest 0.001$^\circ$ (comparable to the typical seeing expected from the all-sky surveys), leaving us with 8003 unique coordinates. We further identify duplicate sources by, first, using the available data in the cross-matched catalogs (e.g., looking for identical radio-counterparts), and then manually inspecting the known properties (e.g., redshift, source type) of all 140 pairs of sources closer than 0.2$^\circ$. Sources suspected to be duplicates are then merged, with their final RA and Dec. preferentially corresponding to their radio counterpart from: RFC $\rightarrow$ 4LAC $\rightarrow$ 3HSP $\rightarrow$ CGRaBS $\rightarrow$ 5BZC. We note that, as seen below, we remove two sources (CGRaBS J1537$-$7154 and RFC~J1911+0458) from the catalog for not being AGN.

\begin{figure*}
    \centering
    \includegraphics[width=18cm]{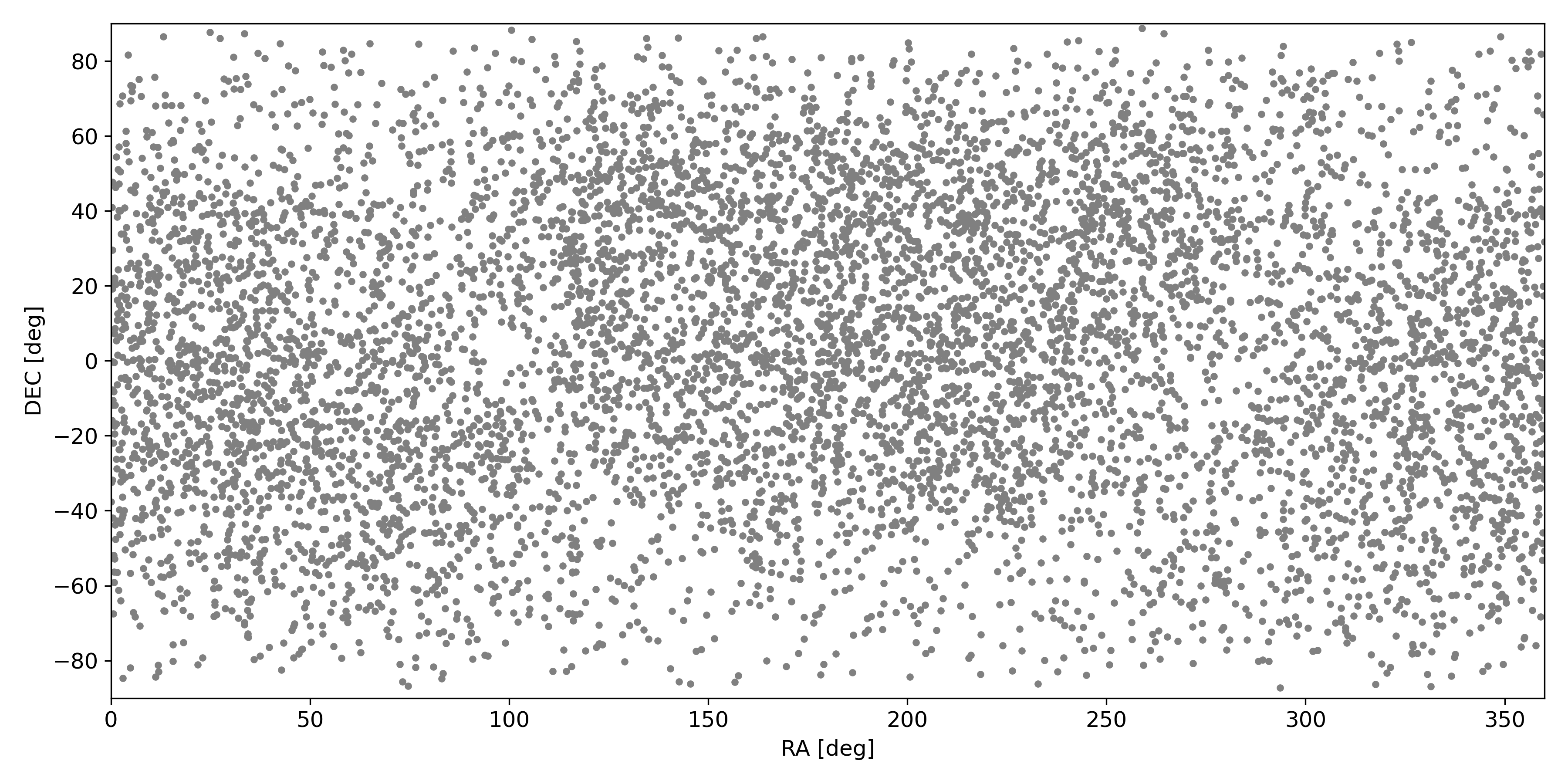}
    \caption{Sky distribution of 7918 sources in the CAZ catalog using the equatorial coordinate system (J2000 epoch).}
    \label{fig_CAZ_skymap}
\end{figure*}

The final CAZ catalog contains 7918 unique AGN and candidates, of which 5111 (64.5\%) are in the RFC catalog, 3225 (40.7\%) are in the statistically complete, flux-limited RFC subsample\footnote{The initial RFC flux-limited subsample has 3236 sources meeting $S_\text{X-band}^{\text{VLBI}} \geq 150$~mJy; however, ten of them have neighboring sources closer than 0.001$^\circ$ and one of them is not an AGN. These 11 are removed from the CAZ catalog.} referred to as RFC$^\dag$, 3814 (48.2\%) are in the 4LAC-DR3 catalog, 2013 (25.4\%) are in the 3HSP catalog, 1625 (20.5\%) are in the CGRaBS sample, and 3561 (45.0\%) are in the 5BZC catalog. Moreover, at least 696 (8.8\%) of these 7918 sources have been a monitoring target in one of the major blazar monitoring programs\footnote{\href{https://www.cv.nrao.edu/MOJAVE/blazarlist.html}{https://www.cv.nrao.edu/MOJAVE/blazarlist.html}}. The sky locations of the sources in the CAZ catalog are shown in Fig. \ref{fig_CAZ_skymap}.

For as many of the 7918 sources in the CAZ catalog as possible, we collect source types, redshifts, SED peak frequencies, and radio variability Doppler factors mainly from the above catalogs. We also extract X-ray flux densities when possible. Additionally, we extract redshift and source type information from the Set of Identifications, Measurements, and Bibliography for Astronomical Data (SIMBAD\footnote{\href{https://simbad.cds.unistra.fr/simbad/}{https://simbad.cds.unistra.fr/simbad/}}) for as many sources as possible.

The above catalogs and SIMBAD have somewhat different approaches of classifying their sources. Therefore, we devised our own AGN classes to unify the distinct classifications. Our blazar-focused AGN classification scheme is shown below:
\begin{itemize}
    \item ``Q'': FSRQs and FSRQ candidates
    \item ``B'': BLLs and BLL candidates
    \item ``G'': Host-galaxy-dominated BLLs\footnote{This is a 5BZC exclusive category which we do not merge with any other blazar category.}
    \item ``U'': Blazars of unknown type and blazar candidates
    \item ``A'': Confirmed non-blazar AGN\footnote{This includes radio galaxies, Seyfert type galaxies, etc.} and AGN candidates
\end{itemize}

Many sources have classification information from multiple references. We accounted for this by prioritizing the catalogs as 4LAC $\rightarrow$ CGRaBS $\rightarrow$ 3HSP $\rightarrow$ 5BZC $\rightarrow$ SIMBAD. We note that the 3HSP catalog does not have source classification, but all of its sources are identified as HSPs, making them BLLs. If a radio source does not have a classification from any of the catalogs, it is taken as an AGN candidate. The 7918 AGN and candidates in the CAZ catalog fit into our blazar-focused classification as shown below:
\begin{itemize}
    \item Q: 2495 (31.5\%)
    \item B: 2726 (34.4\%)
    \item G: 112 (1.4\%)
    \item U: 1697 (21.4\%)
    \item A: 888 (11.2\%)
\end{itemize}
Therefore, 88.8\% of the 7918 sources in the CAZ catalog are blazars or blazar candidates. We included the 11.2\% non-blazar AGN and AGN candidates because the classifications from different catalogs do not always agree, so it is possible that some of the currently classified non-blazar AGN are actually blazars. Furthermore, by minimally excluding sources, we maintain continuity between the CAZ and its parent catalogs (i.e., 4LAC, 3HSP, CGRaBS, 5BZC, and the statistically complete subsample of the RFC), making future comparative studies between them more straightforward.

\begin{figure}
    \centering
    \includegraphics[width=0.49\textwidth]{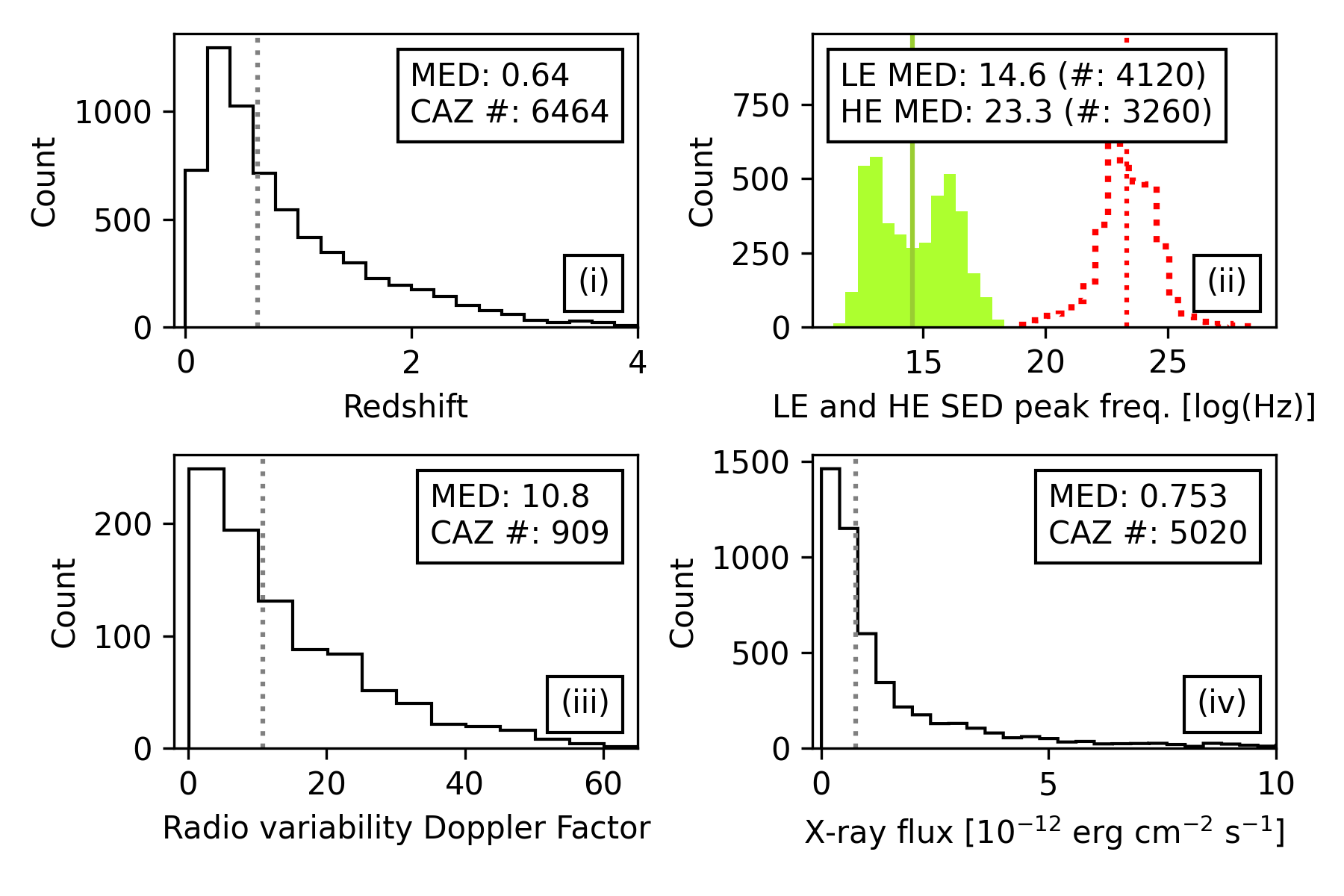}
    \caption{Distributions of (i) redshift, (ii) LE and HE peak frequencies (shown in green and red, respectively), (iii) radio variability Doppler factor, and (iv) median X-ray flux density of sources in the CAZ catalog. ``MED'' refers to median, and ``CAZ \#'' refers to the number of available parameters.}
    \label{fig_CAZ_basic_dist}
\end{figure}

There are a total of 6464 redshift estimates for the 7918 sources (i.e., 82.7\% of the CAZ catalog). These are primarily collected from the catalogs mentioned above as well as the Optical Characteristics of Astrometric Radio Sources (OCARS) catalog, which contains both spectroscopic and photometric redshift estimates (\citealt{Malkin2018_OCARS_cat}). Lastly, we also search SIMBAD for redshift estimates. Since the reliability of these estimates differ and there are often mismatching estimates, we select the final redshift for each source using the following priority order: 4LAC $\rightarrow$ CGRaBS $\rightarrow$ 5BZC $\rightarrow$ SIMBAD $\rightarrow$ OCARS $\rightarrow$ 3HSP. This priority order is identical to the classification priority order with the exception of 3HSP. The 4LAC catalog has the highest priority because all of its redshift estimates are spectroscopic and their reliability has been
checked rigorously, while the 3HSP catalog has the lowest priority because it primarily reports photometric redshifts. The distribution of the 6464 redshift estimates is shown in Fig. \ref{fig_CAZ_basic_dist} (i).

We have a total of 4120 (52.0\% of 7918) low-energy (LE; i.e., synchrotron) peak frequencies from the 4LAC (2825 of 4120) and 3HSP (1295 of 4120) catalogs, where the former value is prioritized in case of duplicates. In addition, there are a total of 3260 (41.2\% of 7918) high-energy (HE) peak frequencies from the 4LAC catalog. The distributions of these LE and HE peak frequencies are shown in Fig. \ref{fig_CAZ_basic_dist} (ii). Based on these LE peak frequencies, we consequently identify 1699 sources (41.2\% of 4120) as LSPs, 536 (13.0\%) as ISPs, 1619 (39.3\%) as HSPs, and 266 (6.5\%) as EHSPs using the synchrotron peak frequency thresholds defined in Sect. \ref{sec_intro}.

Moreover, we collected 909 (11.5\% of 7918) radio variability Doppler factors for the sources in the CAZ catalog. The majority of these (875 of 909) are from \cite{liodakis2018_VarDopplerFac}, and the remaining 34 are calculated in this study using previously unavailable redshift information via Eq. 2 of \cite{liodakis2018_VarDopplerFac}. The distribution of these radio variability Doppler factors is shown in Fig. \ref{fig_CAZ_basic_dist} (iii).

Lastly, we obtained a median X-ray flux density for 5020 (63.4\% of 7918) sources in the CAZ catalog. To do this, in our companion paper Paggi et al. (in prep.), we collect X-ray light curves for as many of the CAZ sources as possible using archival data from the space-borne \textit{Neil Gehrels Swift} Observatory (\textit{Swift}-XRT\footnote{\href{https://heasarc.gsfc.nasa.gov/}{https://heasarc.gsfc.nasa.gov/}}; \citealt{Gehrels2004_swift}), \textit{Chandra} X-ray Observatory\footnote{\href{http://cda.harvard.edu/chaser}{http://cda.harvard.edu/chaser}} (e.g., \citealt{Weisskopf2002_chandra}), and X-ray Multi-Mirror Mission (XMM-\textit{Newton}\footnote{\href{http://nxsa.esac.esa.int/nxsa-web}{http://nxsa.esac.esa.int/nxsa-web}}; \citealt{Jansen2001_XMM}). We then calculated the median of the detected (3$\sigma$) X-ray flux densities in these light curves. For 5020 sources in the CAZ catalog, the X-ray light curve has at least one detection. This leads to 5020 median flux density measurements, whose distribution is shown in Fig. \ref{fig_CAZ_basic_dist} (iv). For 734 (9.3\% of 7918) sources in the CAZ catalog, we only find upper limits on the X-ray flux density. More details about the X-ray data extraction is given in Paggi et al. (in prep.), where we also provide the X-ray light curves and spectra of these sources. In that paper, we also construct an X-ray-selected, statistically complete sample of blazars and perform a number of complementary blazar-population analyses.

\section{CAZ light curves} \label{sec_lc_extraction}
Here we describe the process of obtaining optical light curves for the 7918 sources in the blazar-dominated CAZ catalog. In short, we constructed decade-long and daily cadenced optical light curves for as many sources as possible, mainly using three all-sky surveys: CRTS, ATLAS, and ZTF (Sect. \ref{sec_intro}). We additionally obtained 408 white dwarf light curves (referred to as ``WD'' CAZ light curves), which are used as non-variable control light curves in the ensuing variability analysis (Sect. \ref{sec_characterizing_variability}). The CAZ light curves have data until the end of 2023, i.e., Modified Julian Date (MJD) of 60310. The final CAZ light curves are provided as electronic tables.

\subsection{Extracting raw data} \label{sec_lc_raw}
We collected data from the CRTS\footnote{\href{http://crts.caltech.edu/}{http://crts.caltech.edu/}} all-sky survey, which officially ran from 2007 to 2016 (\citealt{drake2009}). This survey is obtained without a filter (i.e., closest to the standard V filter with $\lambda_\mathrm{eff}=551$~nm) via three telescopes: two located in Arizona, with limiting magnitudes of 19.5 and 21.5, and one in Australia with a limiting magnitude of 19.0. Subsequently, we collected data from the ATLAS\footnote{\href{https://fallingstar-data.com/}{https://fallingstar-data.com/}} all-sky survey, running since 2016 (\citealt{tonry2018}). It primarily observes in filters cyan (c; $\lambda_\mathrm{eff}=533$~nm) and orange (o; $\lambda_\mathrm{eff}=578$~nm) via four telescopes: two located in Hawaii, one in Chile, and one in South Africa. It also occasionally observes in g ($\lambda_\mathrm{eff}=475$~nm), r ($\lambda_\mathrm{eff}=637$~nm), and i ($\lambda_\mathrm{eff}=783$~nm) filters. The typical limiting magnitude of ATLAS has been $\sim$19 since 2019 (prior to 2019 it was dimmer; \citealt{smith2020}). We then collected data from the ZTF\footnote{\href{https://ztf.caltech.edu/}{https://ztf.caltech.edu/}} all-sky survey, running since 2018 (\citealt{bellm2019}). We used its 20$^\mathrm{th}$ data release, containing data between March 2018 and Jan 2024. ZTF observes using the Palomar 48" Schmidt telescope located in California in three filters: g, r, and i. The limiting magnitude for g and r is $\sim$21.5, and for i $\sim$21.0. In addition to these all-sky survey light curves, we included data from our Tuorla blazar monitoring program\footnote{\href{https://tuorlablazar.utu.fi/}{https://tuorlablazar.utu.fi/}} and the Katzman Automatic Imaging Telescope (KAIT\footnote{\href{https://w.astro.berkeley.edu/bait/kait.html}{https://w.astro.berkeley.edu/bait/kait.html}}) monitoring program. As part of the Tuorla program we have been monitoring 205 blazars in the R-band ($\lambda_\mathrm{eff}=658$~nm) since 2002 via various telescopes (\citealt{nilsson2018}), and in this paper we publish our post-2013 data for the first time. KAIT has been searching for supernovae in the V filter since 1997 \citep{filippenko2001} and has ended up monitoring 155 blazars serendipitously (e.g., \citealt{Cohen2014_KAIT_blazars, liodakis2018}).

While for the two dedicated monitoring programs, source-based light curves in flux density versus MJD are available, in the case of the three all-sky surveys forced photometry at each source coordinate (within $\pm$0.001$^\circ$) is needed to obtain the light curves. If only magnitudes are available, we convert them to flux densities using $S=S_0 \cdot 10^{-0.4m}$ where $S$, $S_0$, and $m$ are flux density, flux density zero-point, and apparent magnitude, respectively. We use the following flux density zero-points for each band: $S_{0,\mathrm{V}}=3640$~Jy, $S_{0,\mathrm{R}}=3080$~Jy (\citealt{mead1990}), and $S_{0,\mathrm{g,r,i}}=3631$~Jy (referring to the monochromatically defined g, r, and i AB-magnitudes; \citealt{tonry2012}).

The light curve quality across these all-sky surveys and dedicated monitoring programs differs greatly. While the latter are generally of higher quality due to supervised data collection and reduction, the all-sky surveys (especially ATLAS) suffer from outliers and low signal-to-noise ratio (S/N) observations due to automatic forced-photometry and data reduction algorithms. Manual inspection of the raw light curves reveals that ATLAS light curves exhibit unrealistically bright data points (e.g., the brightest ATLAS data point is around $-$38~magnitudes). Since the brightest raw data points from all other surveys and programs are $\gg$5~magnitudes, we exclude all ATLAS raw data points brighter than 5~magnitudes, corresponding to $\sim$0.6\% of $\sim$18.9M ATLAS data points. Subsequently, we impose S/N~$>$~1 on all extracted light curves, which excludes$\sim$0.01\% of $\sim$1.6M CRTS raw data points, $\sim$11.7\% of $\sim$18.9M ATLAS, $\sim$0.01\% of $\sim$6.3M ZTF, $\sim$0.01\% of $\sim$37k KAIT, and none of $\sim$45k Tuorla data points. We note that these estimates include the 408 WD light curves in addition to the 7918 AGN light curves. More strict detection criteria (S/N~$\ge$~2) are imposed after cleaning and nightly binning the light curves (see Sect. \ref{sec_clean_outliers}).

The majority of the CRTS raw data is taken from the online repository\footnote{\href{http://nunuku.caltech.edu/cgi-bin/getmulticonedb_release2.cgi}{http://nunuku.caltech.edu/cgi-bin/getmulticonedb\_release2.cgi}}. In addition to this, we have access to extended CRTS light curves (by $\sim$1~yr for some sources; as early as mid-2005), which we add onto their public data. The ATLAS\footnote{\href{https://fallingstar-data.com/forcedphot/apiguide/}{https://fallingstar-data.com/forcedphot/apiguide/}} and ZTF\footnote{\href{https://irsa.ipac.caltech.edu/docs/program_interface/ztf_api.html}{https://irsa.ipac.caltech.edu/docs/program\_interface/ztf\_api.html}} raw data are also extracted using their respective public repositories. Overall, 5185 (65.5\% of 7918) sources have at least some data points in their extracted CRTS light curves, all 7918 in their ATLAS light curves, and 5607 (70.8\% of 7918) in their ZTF light curves. From the Tuorla sample, there are 205 blazar light curves present, while from the KAIT sample there are 155.

\subsection{Cleaning outliers and nightly binning} \label{sec_clean_outliers}
While blazars are highly variable, even on timescales of minutes, the above surveys and monitoring programs cannot detect such fast, intranight variability reliably. Therefore, we assume that all intranight variability is due to noise and bin all the flux densities on a nightly basis. Before binning, we first ensure that a filter has at least three data points across one source, then we use the criteria introduced by \cite{Hovatta2014_opt_vs_gamma_variability} to identify and remove outliers: (1) we find the mean and standard deviation of the intranight magnitude values, (2) if standard deviation is greater than 0.5 magnitudes, we exclude the entire night, otherwise we select the brightest and the dimmest data points, (3) if either the brightest or dimmest data is different from the mean by more than 0.8 magnitudes, we exclude the more extreme of the two, and (4) we repeat the process until there are no intranight magnitude measurements with a standard deviation $>$0.5.

Subsequently, we calculated a binned flux density for each night (with $S_i \pm \sigma_i$ flux densities and errors, where $i=1,\cdots,N$) using inverse-variance weighted averaging:

\begin{equation} \label{eqn_wAvg}
    \overline{S} = \frac{\sum_{i=1}^{N} S_i w_i}{\sum_{i=1}^{N} w_i}
\end{equation}
\begin{equation} \label{eqn_wAvg_error_standard}
    \sigma_{\overline{S}} = \sqrt{\frac{1}{\sum_{i=1}^{N} w_i}},
\end{equation}
where $w_i=\sigma_i^{-2}$ and $\overline{S}$ is the binned flux density, with $\sigma_{\overline{S}}$ being its error. To ensure the robustness of the nightly binned errors ($\sigma_\mathrm{night}$), we added 1.2\% of $\overline{S}$ to $\sigma_{\overline{S}}$ in quadrature as a flat-fielding factor via $\sigma_\mathrm{night} = [ \sigma_{\overline{S}}^2 + (0.012 \times \overline{S})^2 ]^{1/2}$ (e.g., \citealt{nilsson2018}). We excluded all nights with S/N~$=\overline{S}/\sigma_\mathrm{night} < 2$. If there was only one data point per night, we required S/N~$ \ge 3$. The above criteria exclude $\sim$3.0\% of $\sim$1.6M CRTS data points (after implementing the basic exclusion criteria of Sect. \ref{sec_lc_raw}), $\sim$7.8\% of $\sim$16.6M ATLAS, $\sim$2.4\% of 6.3M ZTF, $\sim$11.8\% of $\sim$37k KAIT, and $\sim$2.7\% of $\sim$45k Tuorla.

\subsection{Merging all filters} \label{sec_merge_all_filters}

\begin{figure*}
    \centering
    \includegraphics[width=18cm]{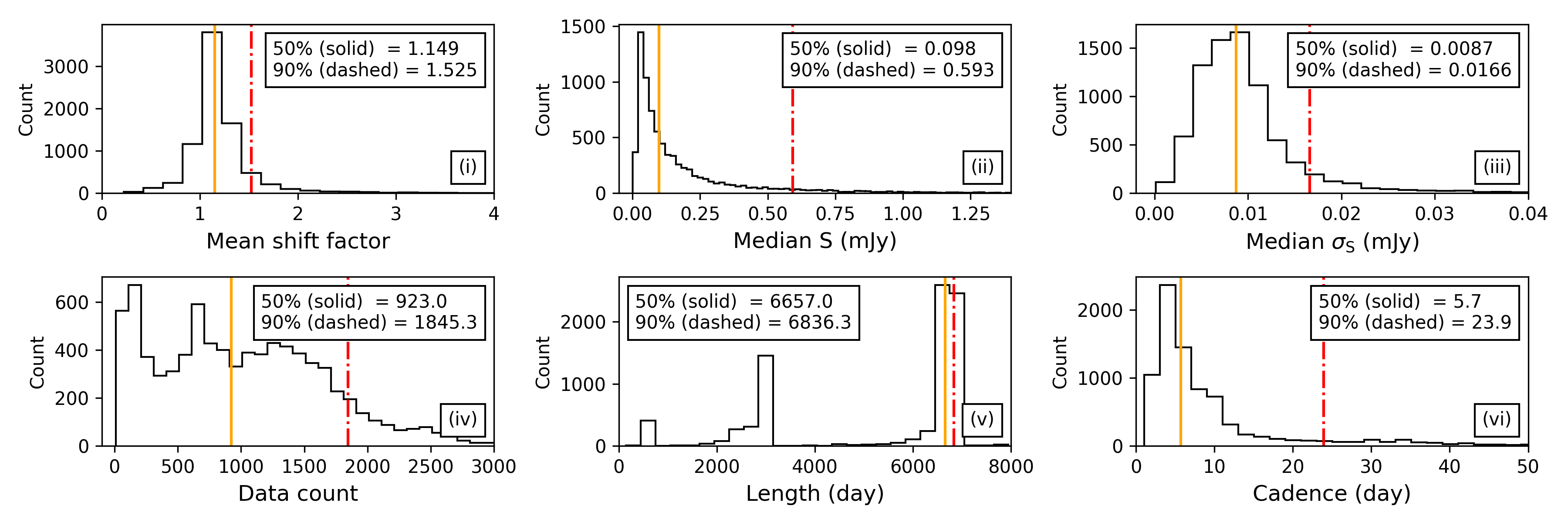}
    \caption{Global distributions of 8130 merged CAZ light curves (7722 of AGN and 408 of WDs) with $\ge$10 data points and $\ge$30~d duration. The panels are as follows: (i) Mean shift factor across all filters per light curve. (ii) Median flux density per light curve. (iii) Median flux density error per light curve. (iv) Data count per light curve. (v) Length of each light curve. (vi) Cadence per light curve calculated by dividing the length of each light curve by its data count. The vertical solid orange line shows the median, and the dashed red line shows the 90$^\mathrm{th}$ percentile.}
    \label{fig_merged_LC_stats}
\end{figure*}

All the above treatments were applied to data points within one filter of a given survey or monitoring programs. There are a total of 11 unique survey and filter combinations: one from CRTS, denoted as: [C]V; one from KAIT, [K]V; one from Tuorla, [T]R; five from ATLAS, [A]o, [A]c, [A]r, [A]g, and [A]i; and three from ZTF, [Z]r, [Z]g, and [Z]i. We merge these in order to maximize the cadence of the final CAZ light curves. Merging is ideally done via empirically well-established filter transformation relations when the spectral shape of a source is known. Unfortunately, we neither have the filter transformation relations, nor the source-specific spectral shapes. Another possibility is to use simultaneous multicolor data to calculate a filter transformation relation\footnote{For example, to convert g and r AB-magnitudes to Cousins V-band magnitude, the linear relationships between $m_\mathrm{g}-m_\mathrm{r}$ versus g and $m_\mathrm{g}-m_\mathrm{r}$ versus r are needed, respectively (e.g., \citealt{tonry2012}).}. However, such relations must be found uniquely for each source, because blazar emission can be strongly chromatic with a wide range of spectral shapes from one source to another and show varying degrees of chromaticity at different brightness levels (i.e., bluer-when-brighter and redder-when-brighter trends; e.g., \citealt{Rani2010_blz_chromaticity, Bonning2012_blz_chromaticity, Sandrinelli2014_blz_chromaticity, Zhang2015_chromaticity, Meng2018_blz_chromaticity, OteroSantos2022_blz_optical_components}). Unfortunately, due to the lack of simultaneous multicolor data, we cannot do this either. Thus, we choose not to perform filter transformations but translate the filters in flux density space by multiplying their flux densities and errors\footnote{Multiplying the error bar ensures that the S/N of the translated data point remains constant.} with a shift factor, determined automatically on a filter-by-filter and source-by-source basis. This approach reliably eliminates overall flux density discrepancies arising due to a constant degree of chromaticity (e.g., when a filter is brighter than another by a constant factor on average). However, it cannot reliably account for flux density discrepancies which arise when the degree of chromaticity changes over time (e.g., due to bluer-when-brighter or redder-when-brighter trends). Nevertheless, translating the filters seems to preserve the variability information of most CAZ light curves rather well, making it suitable for our purposes.

The shift factor for a filter and survey pair (two of the aforementioned 11, e.g., [S$_1$]f$_i$ and [S$_2$]f$_j$; hereafter, simply referred to as a filter) was determined via one of four techniques, in decreasing order of priority:
\begin{itemize}
    \item the average ratio of at least five simultaneous (within $\pm$1~d) flux densities,
    \item same as above, except $\pm$2~d is used to define simultaneous flux densities,
    \item the ratio of the average flux density of one filter to the other within their overlapping time window, or
    \item same as above, except the overlapping time window is calculated with a $\pm$50~d extension.
\end{itemize}
In the last two techniques, to calculate the overlapping flux density averages, there must exist at least five overlapping data points for each filter. Otherwise, they are not shifted at all. We began by listing all possible filter pairs for a light curve. We then checked if any of them can be shifted using the highest priority technique. If yes, one is shifted onto the other, and they are merged into one: [S$_1$]f$_i$+[S$_2$]f$_j$. A merged filter was then treated as an independent filter and sought to be merged with other filters. If a shift technique does not result in any mergers, we applied the next shift technique in the priority list. This is repeated until either all filters in the light curve were merged into one or when the last shift technique cannot result in any mergers. Globally, 86.4\%, 7.2\%, 5.5\%, and 0.9\% of the shifts are done using the four techniques in decreasing order of priority, respectively. Whenever a filter cannot be shifted, we reported its original flux densities. Finally, the reference filter (i.e., the filter onto which all other filters are shifted) for each light curve is chosen based on data availability. In decreasing order of priority, we chose the reference filter to be [Z]r, [Z]g, [Z]i, [C]V, [T]R, [K]V, [A]o, [A]c, [A]r, [A]g, and [A]i. We report the final shift factor for each filter in the electronic table. A shift factor of 1 implies that the filter is not shifted (e.g., the reference filter). Notably, all shifted filters can be shifted back to their original flux density level by dividing their flux densities and flux density errors with the final shift factor we provide in the electronic table.

In Fig. \ref{fig_merged_LC_stats}, we show the global distribution of average shift factor across filters, median flux density, median flux density error, data count, length, and cadence per light curve for the merged light curves with $\ge$10 data points and $\ge$30~d duration (corresponding to 7722 sources in the CAZ catalog and 408 WDs). The merged light curves typically have a total of $\sim$900 data points spanning $\sim$18~years in length, translating to an average weekly cadence. However, we emphasize that deviations from these typical values often happen, and these global averages do not visualize potential source-by-source caveats (e.g., large seasonal gaps, noise-like behavior, outliers). Notably, combining three all-sky surveys creates three distinct global peaks in the distributions of the data count and length.

\begin{figure*}
    \centering
    \includegraphics[width=18cm]{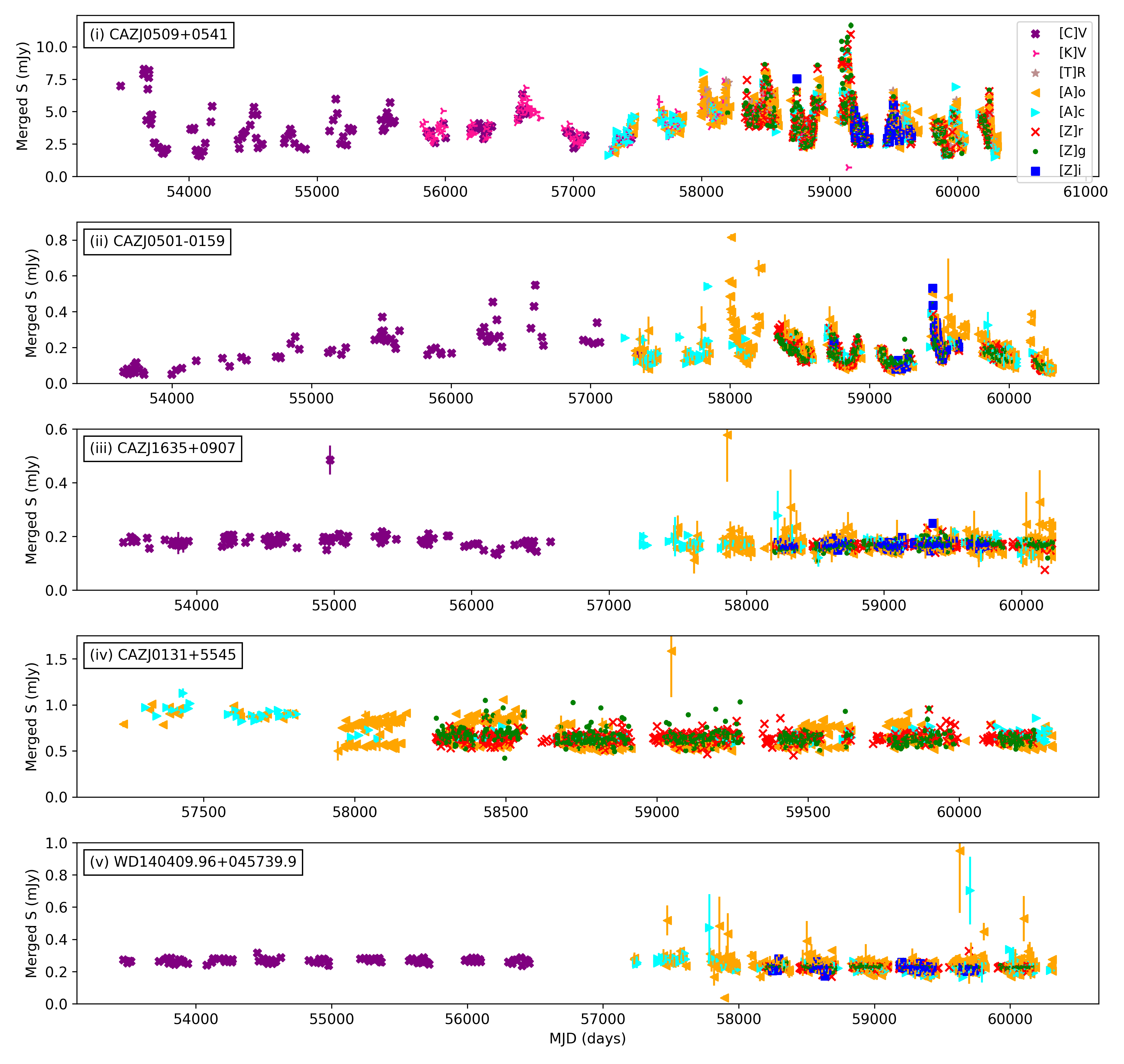}
    \caption{Example merged light curves. The name of the source is given on the top-left corner of each panel. We note that some light curves are of much higher quality than others (e.g., plot i compared to iv; the former is of the well-known, highly variable TXS~0506+056, while the latter is of a host-galaxy-dominated source). A zoomed-in version of these light curves (including identified periods of enhanced emission and flares) is given in Fig. \ref{fig_eg_LCs_eg_flares}.}
    \label{fig_eg_LCs}
\end{figure*}

\begin{figure*}
    \centering
    \includegraphics[width=18cm]{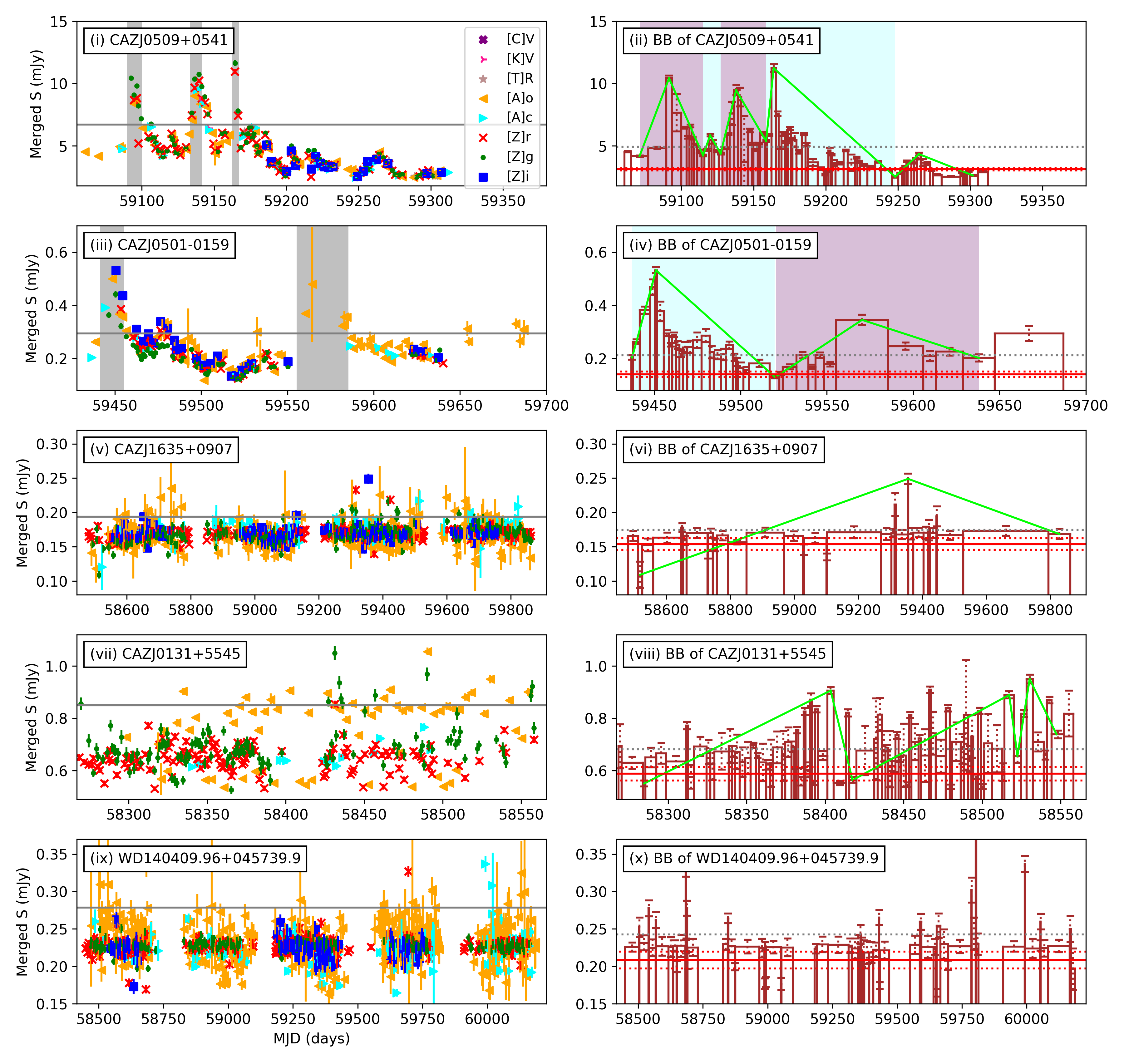}
    \caption{Example zoomed-in light curves with BB95 periods (Sect. \ref{sec_identifying_critically_bright_BBs}) and BBHOP flares (Sect. \ref{sec_identifying_flares}) identified. The merged light curves on the left are zoomed-in versions of those shown in Fig. \ref{fig_eg_LCs}. The light curves on the right are their corresponding BB light curves. The gray vertical area on the left shows BB95 periods, which are BBs whose flux density is larger than $S_{95\%}$ (shown as the gray horizontal line in the left). BBHOP flares are shown on the right using green lines, connecting their start to the peak and then to the end BB. Prominent BBHOP flares are further distinguished with a purple or cyan background. The horizontal solid red line in the right panels represents the reference flux density for determining the global amplitude $\mathcal{G}$ (i.e., $S_{25\%}$; the dotted lines represent its error). The gray horizontal dotted line in the right plots represents $S_{75\%}$, which is needed when determining prominent BBHOP flares. We note that CAZJ1635+0907 and CAZJ0131+5545 were determined to be non-variable, and WD140409.96+045739.9 is non-variable by definition. Thus, we did not identify BB95 and prominent BBHOP flares in their light curves.}
    \label{fig_eg_LCs_eg_flares}
\end{figure*}

In Fig. \ref{fig_eg_LCs} we show five example merged light curves: from top to bottom, the light curves of CAZJ0509+0541 (TXS~0506+056; a highly variable blazar), CAZJ0501$-$0159 (another variable blazar), CAZJ1635+0907 (a non-variable blazar), CAZJ0131+5545 (a host-galaxy-dominated blazar), and WD140409.96+045739.9 (a white dwarf) are plotted. In Fig. \ref{fig_eg_LCs_eg_flares} we present zoomed-in versions of these light curves where we also show a few of their identified periods of enhanced emission and flares (see Sect. \ref{sec_characterizing_variability}). As evident, some of the light curves are of much lower quality than others. While the merging process can introduce some artifacts, the most notable problems stem from the raw data. For example, as seen at around MJD~58000 in panel (iv) of Fig. \ref{fig_eg_LCs} (also, see panel vii of Fig. \ref{fig_eg_LCs_eg_flares}), the light curve of the host-galaxy-dominated source CAZJ0131+5545 shows two states of constant flux. This is due to the raw ATLAS light curves, which often show such problematic behavior around that epoch in host-galaxy-dominated light curves. While we do not know their cause, we attempt to mitigate their effect on the results of this study at various steps (see, e.g., Sect. \ref{sec_variability}). In a future iteration of this work, we aim to use specialized tools designed to mitigate issues with ATLAS forced photometry, such as ATClean (\citealt{Rest2023_ATClean}), of which we were not aware at the time of constructing the CAZ light curves of this work.

\section{Characterizing variability and flaring behavior} \label{sec_characterizing_variability}
In this section we describe how we quantify the overall variability of the merged CAZ light curves using fractional variability (e.g., \citealt{sokolovsky2016}; see Sect. \ref{sec_variability}), and how we identify periods of enhanced emission and flaring periods in them utilizing the Bayesian block method (\citealt{scargle2013}; see Sect. \ref{sec_identifying_critically_bright_BBs} and \ref{sec_identifying_flares}). We provide the fractional variability measures and flaring periods of the CAZ light curves in the electronic tables.

Since complex color relations and differences in observational properties of the filters (e.g., survey sensitivities, aperture sizes, typical error sizes, cadence) introduce variability upon merging, we control for such non-intrinsic variability using WDs. While the flux density of WDs can exhibit rather small intrinsic variability (on the order of a few percent of their average flux density; e.g., \citealt{Bell2025_WD_var}), they are non-variable in comparison to blazars.

\subsection{Quantifying variability} \label{sec_variability}
Fractional variability ($F_\mathrm{var}$) is a commonly used variability measure for time series data, especially for AGN and blazar light curves (e.g., \citealt{Gliozzi2003_Fvar, MAGIC2024_mrk501_Fvar}). It quantifies the average deviation of the data in excess of the observational errors. For a sequence of $S_i$ flux densities, where $i=1, \dots, N$, it is defined as

\begin{equation} \label{eqn_Fvar}
    F_\mathrm{var} = \sqrt{\frac{\left[ \frac{1}{N-1} \sum_{i=1}^N (S_i-\overline{S})^2 \right] - \overline{\sigma^2}}{\overline{S}^{~2}}},
\end{equation}
and its error is calculated via
\begin{equation} \label{eqn_FvarErr}
    \Delta F_\mathrm{var} = \sqrt{{F_\mathrm{var}}^2 + \sqrt{\left(\sqrt{\frac{2}{N}} \cdot \frac{\overline{\sigma^2}}{\overline{S}^{~2}}\right)^2+\left(\sqrt{\frac{\overline{\sigma^2}}{N}} \cdot \frac{2 F_\mathrm{var}}{\overline{S}}\right)^2}} - F_\mathrm{var},
\end{equation}
where $\overline{S}=\frac{1}{N} \sum_{i=1}^{N}S_i$ and $\overline{\sigma^2}=\frac{1}{N} \sum_{i=1}^{N}\sigma^2_i$ are the sample mean and mean square error, respectively (\citealt{Vaughan2003_Fvar, Poutanen2008_Fvar_err}).

\begin{figure*}
    \centering
    \includegraphics[width=18cm]{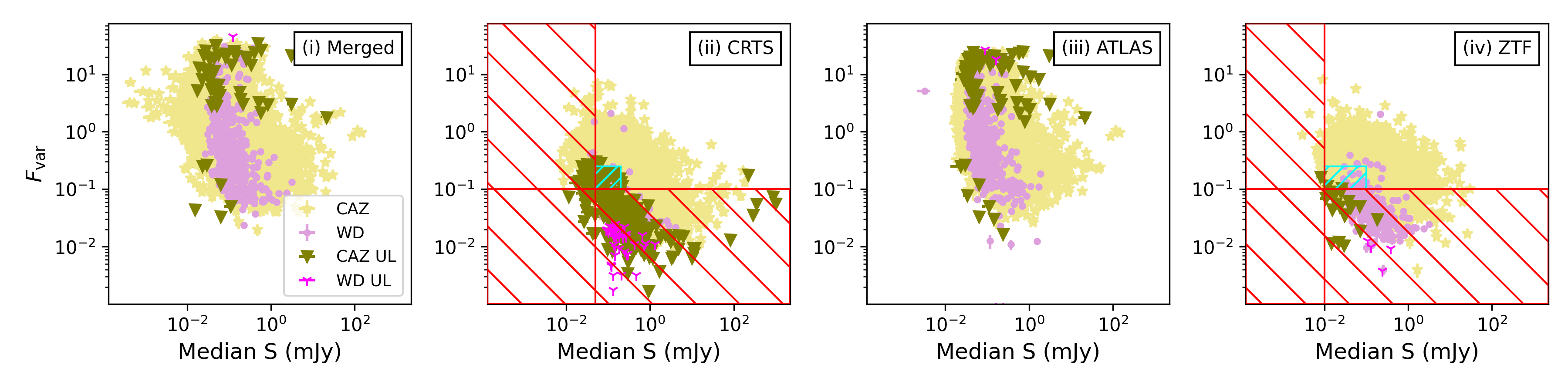}
    \caption{Fractional variability ($F_\mathrm{var}$) versus median flux density plots. The panels are as follows: (i) Merged light curves. (ii) CRTS only. (iii) ATLAS only. (iv) ZTF only. The term ``UL'' refers to $3\sigma$ upper limits of $F_\mathrm{var}$. The red hatched areas in (ii) and (iv) represent the low flux density and low variability regimes to determine non-variable light curves. The cyan hatched box is typically occupied by WDs and is another non-variable regime.}
    \label{fig_Fvar_v_flux}
\end{figure*}

In Fig. \ref{fig_Fvar_v_flux}, we show $F_\mathrm{var}$ of the merged CAZ, CRTS-, ATLAS-, and ZTF-only blazar and WD light curves against their median flux densities. The merged WDs appear to show the same variability as the blazar population, which is unexpected. This problematic behavior is also seen when ATLAS-only merged filters are considered. On the other hand, in both CRTS- and ZTF-only plots, the WDs cluster at smaller variability measures in comparison to majority of the blazar population. This suggests that ATLAS light curves are prone to noise and, since they typically have more data points than the other surveys, their inclusion contaminates the merged CAZ light curves. 

In addition to $F_\mathrm{var}$, we explore quantifying the variability of the merged (CAZ, CRTS-, ATLAS-, and ZTF-only) light curves using three other variability measures: interquartile range (IQR; e.g., \citealt{sokolovsky2016}), inverse \cite{vonNeumann1941_eta} ratio ($1/\eta$; e.g., \citealt{Shin2009_one_over_eta}), and excursions ($E_x$; \citealt{sokolovsky2016}). We describe these in more detail and plot their variability estimate against median flux density in Appendix \ref{appendix_other_var}. While the ATLAS WD scatter is slightly more contained in IQR as compared to $F_\mathrm{var}$, it appears to show a problematic more-variable-when-brighter trend. With $1/\eta$, the ATLAS WD scatter is well contained; however, all surveys exhibit the more-variable-when-brighter behavior. While $E_x$ does not show such behavior, similar to $F_\mathrm{var}$, it suffers from a large scatter in the ATLAS WDs. Therefore, as $F_\mathrm{var}$ is a well-established variability measure for blazars (\citealt{Schleicher2019_Fvar_use}), we opt to use it as the primary metric to quantify the overall variability of the CAZ light curves. However, due to the issues introduced by ATLAS, we only use the $F_\mathrm{var}$ of CRTS- and ZTF-only merged light curves to define global variability criteria.

We followed the methodology of \cite{Hovatta2014_opt_vs_gamma_variability} to define non-variability criteria in Eqn. \ref{eqn_Fvar_variability_cond}. CRTS- or ZTF-only merged light curves with a flux density lower than $S_\mathrm{low}$ or $F_\mathrm{var}$ lower than $F_\mathrm{var,low}$ are considered non-variable. Additionally, light curves that fall within a box-like region defined by $S_\mathrm{box}$ and $F_\mathrm{var,box}$ are considered non-variable for being WD-like. In other words, if either of the following conditions holds, the CRTS- and ZTF-only light curves are deemed non-variable:

\begin{equation} \label{eqn_Fvar_variability_cond}
    \begin{cases}
        \tilde{S} < S_\mathrm{low} & \, \\
        F_\mathrm{var} < F_\mathrm{var,low} & \, \\
        (\tilde{S} < S_\mathrm{box})~\mathrm{AND}~(F_\mathrm{var} < F_\mathrm{var,box}) & \,,
    \end{cases}
\end{equation}
where $\tilde{S}$ refers to the median flux density. In the CRTS-only case, $S_\mathrm{low}=0.05~\mathrm{mJy}$, $F_\mathrm{var,low}=0.10$, $S_\mathrm{box}=0.20~\mathrm{mJy}$, and $F_\mathrm{var,box}=0.25$. In the ZTF-only case, $S_\mathrm{low}=0.01~\mathrm{mJy}$, $F_\mathrm{var,low}=0.10$, $S_\mathrm{box}=0.10~\mathrm{mJy}$, and $F_\mathrm{var,box}=0.25$. These non-variable regimes are optimized by eye and are marked on the CRTS- and ZTF-only plots in Fig. \ref{fig_Fvar_v_flux}.

The criteria of Eq. \ref{eqn_Fvar_variability_cond} only determine if the CRTS or ZTF portions of a CAZ light curve are variable or not. To confidently identify a CAZ light curve as variable, we required either CRTS- or ZTF-only portions to be variable with $F_\mathrm{var}>0.30$. If $F_\mathrm{var}\leq 0.30$, then they both need to be variable simultaneously for a confident variable identification. This extra condition ensures that a slight outlier-induced increase in the $F_\mathrm{var}$ of either CRTS- or ZTF-only portions does not falsely identify the source as variable. Notably, without CRTS or ZTF data, we cannot reliably establish the variability of a CAZ light curve. This means that light curves with only ATLAS data are conservatively assumed to be non-variable. When applied to 7722 CAZ light curves with at least some CRTS or ZTF data, which have $\ge$10 data points and are $\ge$30~d in duration, these criteria identify 2798 (36.2\%) as confidently variable.

\subsection{Identifying periods of enhanced emission} \label{sec_identifying_critically_bright_BBs}
The Bayesian block algorithm (\citealt{scargle2013}) is a widely used technique for characterizing flux variations and identifying flares in blazar light curves (e.g., \citealt{Wagner2022_BBHOP_flips, deJaeger2023_ASSASN_time_lags}). It is especially powerful for systematically characterizing flares in a large number of light curves, due to its ability to adaptively bin a light curve using only one easily quantifiable Bayesian prior. Therefore, in this subsection we use it on the merged CAZ light curves (see Sect. \ref{sec_merge_all_filters}) to identify periods of enhanced emission and, in Sect. \ref{sec_identifying_flares}, flaring periods. The Bayesian blocks corresponding to the CAZ light curves are provided as electronic tables.

We apply the Bayesian block algorithm using its implementation in the Python package \texttt{astropy} (\citealt{Astropy2022_v5}) to 7722 merged CAZ light curves with $\ge$10 data points and $\ge$30~d duration. The algorithm then adaptively determines instances of significant flux density variation, between which lie periods (i.e., ``blocks'') of insignificant variation. We bin the flux density within each Bayesian block (BB). The binned flux densities are obtained via Eq. \ref{eqn_wAvg} and their error via

\begin{equation} \label{eqn_wAvg_error_spread}
\sigma_{\overline{S},\mathrm{BB}} = \max
    \Bigg(
        \tilde{\sigma}_{i} \, , \, 
        \sqrt{\frac{\sum_{i=1}^{N} (S_i - \overline{S})^2 \cdot w_i}{\sum_{i=1}^{N} w_i}} \, \,
    \Bigg),
\end{equation}
where $\tilde{\sigma}_{i}$ is the median flux density error within a BB and, similar to Eq. \ref{eqn_wAvg}, $w_i={\sigma_{i}}^{-2}$ and $\overline{S}$ is the binned flux density for each BB. Equation \ref{eqn_wAvg_error_spread} ensures that $\sigma_{\overline{S},\mathrm{BB}}$ resembles either the typical error or the spread of the flux densities within the BB. Furthermore, during this process, we preserved observational gaps ($>$60~d periods with no data) by leaving them as empty periods in the BB light curve. We also ensured that all BBs are at least 1~d in length by iteratively merging and rebinning shorter BBs.

The BB algorithm operates using only one input parameter: a Bayesian prior called \texttt{ncp\_prior}\footnote{Another parameter \texttt{gamma} is often used as the Bayesian prior, which follows the relation \texttt{ncp\_prior}~=~$-\ln$(\texttt{gamma}).}. As \texttt{ncp\_prior}~$\rightarrow \infty$, the algorithm finds fewer BB; conversely, when \texttt{ncp\_prior}~$\rightarrow 0$, more BBs are found. For time series with ideal Gaussian errors, it is empirically shown that \texttt{ncp\_prior}$_\mathrm{ideal}$~$=1.32 + 0.577 \log_{10}(N)$ where $N$ is the total data count (\citealt{scargle2013}). Unfortunately, this ideal assumption is not applicable to the merged CAZ light curves as they suffer from noise and non-intrinsic variability, especially due to the inclusion of ATLAS data (as shown in Sect. \ref{sec_variability}). 

We minimized the effect of noise by calibrating \texttt{ncp\_prior} using merged WD light curves. This process is described in more detail in Appendix \ref{appendix_BB_calib}. When applying and calibrating the BB algorithm, we treated the CRTS light curves (i.e., the [C]V data) separately. This is because CRTS and other filters overlap minimally (if at all), resulting in two segments with vastly different cadence and, consequently, distinct \texttt{ncp\_prior}$_\mathrm{ideal}$. We find that increasing the \texttt{ncp\_prior}$_\mathrm{ideal}$ of CRTS WD light curves by 0.4 and other WD light curves by 3.4 results in mostly constant BB light curves (see Appendix \ref{appendix_BB_calib}). Thus, we adopt these enlargements to \texttt{ncp\_prior}$_\mathrm{ideal}$ when applying the BB algorithm to the CRTS and non-CRTS light curves of the CAZ catalog. Nevertheless, we caution that the BBs are still affected by extreme flux density outliers.

Due to the minimal overlap between the CRTS and non-CRTS segments, the CAZ light curves are often in two distinct segments that cannot be merged together. The first constitutes the CRTS data, whereas the second segment contains all other data (often merged together). In 5102 (66.1\% of 7722) light curves, the CRTS and non-CRTS segments are separate, while they are merged together in the remaining 2620 (33.9\% of 7722). However, as mentioned above, the BBs calculated for the CRTS and non-CRTS segments need to be distinct regardless of their merge status. We combine the edges of the CRTS and non-CRTS BBs to obtain a single set of BBs in those 2620 light curves where they are merged together.

We identify periods of enhanced optical emission by finding BBs whose flux density is greater than the 95$^\mathrm{th}$ percentile flux density (i.e., $S_{95\%}$), referred to as ``BB95.'' In the 5102 light curves where CRTS and non-CRTS segments are not merged together, we calculate their $S_{95\%}$ separately. For the other 2620, $S_{95\%}$ is calculated globally across the entire CAZ light curves. The BB95 periods are generally reliable and correspond to peaks of the most prominent flaring periods in a light curve. Nonetheless, when applied to CAZ light curves, they suffer from a few caveats: (1) They are easily affected by outliers (e.g., see panel i in Figs. \ref{fig_eg_LCs} and \ref{fig_eg_LCs_eg_flares} where a dim outlier with a flux density of $\sim$0.7~mJy at MJD 59143 causes a BB to be not critically bright); (2) they can be unreliable in poorly shifted light curves; and (3) they exclude most periods of flaring activity since they only focus on BBs brighter than $S_{95\%}$. In Fig. \ref{fig_eg_LCs_eg_flares} left panels, the gray vertical areas show BB95 in a few example light curves.

\begin{figure}
    \centering
    \includegraphics[width=0.49\textwidth]{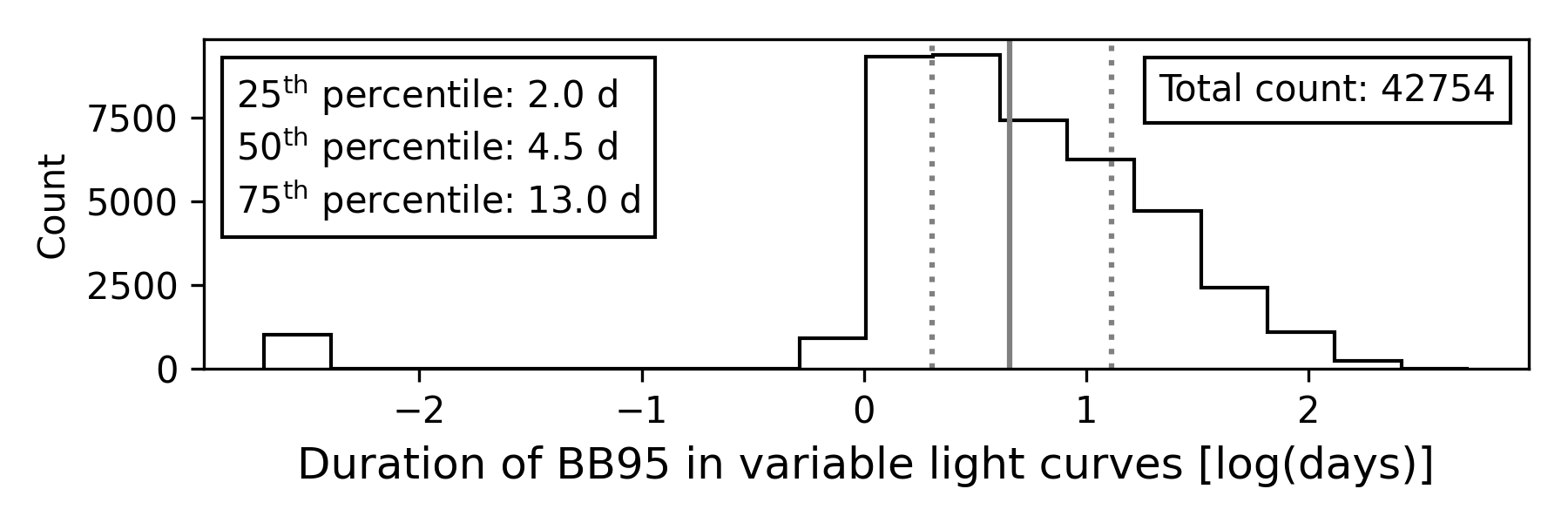}
    \caption{Distribution of the duration of BB95 in variable CAZ light curves. The vertical solid line shows the median of the distribution, and the dotted lines show its 25$^\mathrm{th}$ and 75$^\mathrm{th}$ percentiles.}
    \label{fig_critically_bright_BB_durations_TOP}
\end{figure}

In this study we only focus on BB95 periods of confidently variable light curves (see Sect. \ref{sec_variability}). In total, we find 42754 BB95 across 2769 of 2798 variable sources. The reason for the lack of any BB95 in 29 variable sources is large flux density errors, which cause the binned flux densities to become dimmer than the global 95$^\mathrm{th}$ percentile. In Fig. \ref{fig_critically_bright_BB_durations_TOP} we show the distribution of the BB95 durations, which are typically $\sim$2--13~d. We note that a small peak at duration of $10^{-2.69}$~=~0.002~d arises by construction. If a single data point is preceded and proceeded by observational gaps, we define its BB duration as 2$\times$0.001~d. This is most common in the CRTS epoch.

\subsection{Identifying flaring periods} \label{sec_identifying_flares}
In this section we use BB differently by incorporating up- and down-going trends in flux density to more generally identify flaring periods in merged CAZ light curves. We apply a hopping algorithm (e.g., HOP, \citealt{Eisenstein1998_HOP}) to find ``hills'' and ``valleys'' in the BB light curves. A hill is identified as a BB whose neighboring BBs (both rising and falling) are successively dimmer, and a valley as one whose neighboring BBs are successively brighter. Hills and valleys do not extend over observational gaps (see Sect. \ref{sec_identifying_critically_bright_BBs}) of longer than 60~d. We select all groups of BBs from one valley to the next without going over the gaps. Using this hopping algorithm, each selected group starts on a valley, peaks on a hill, and terminates on a valley. We define each of these valley-hill-valley groups as a ``BBHOP'' flare (e.g., \citealt{Meyer2019_BBHOP}). The final BBHOP flares corresponding to the CAZ light curves are found in the electronic tables.

\begin{figure}
    \centering
    \includegraphics[width=0.49\textwidth]{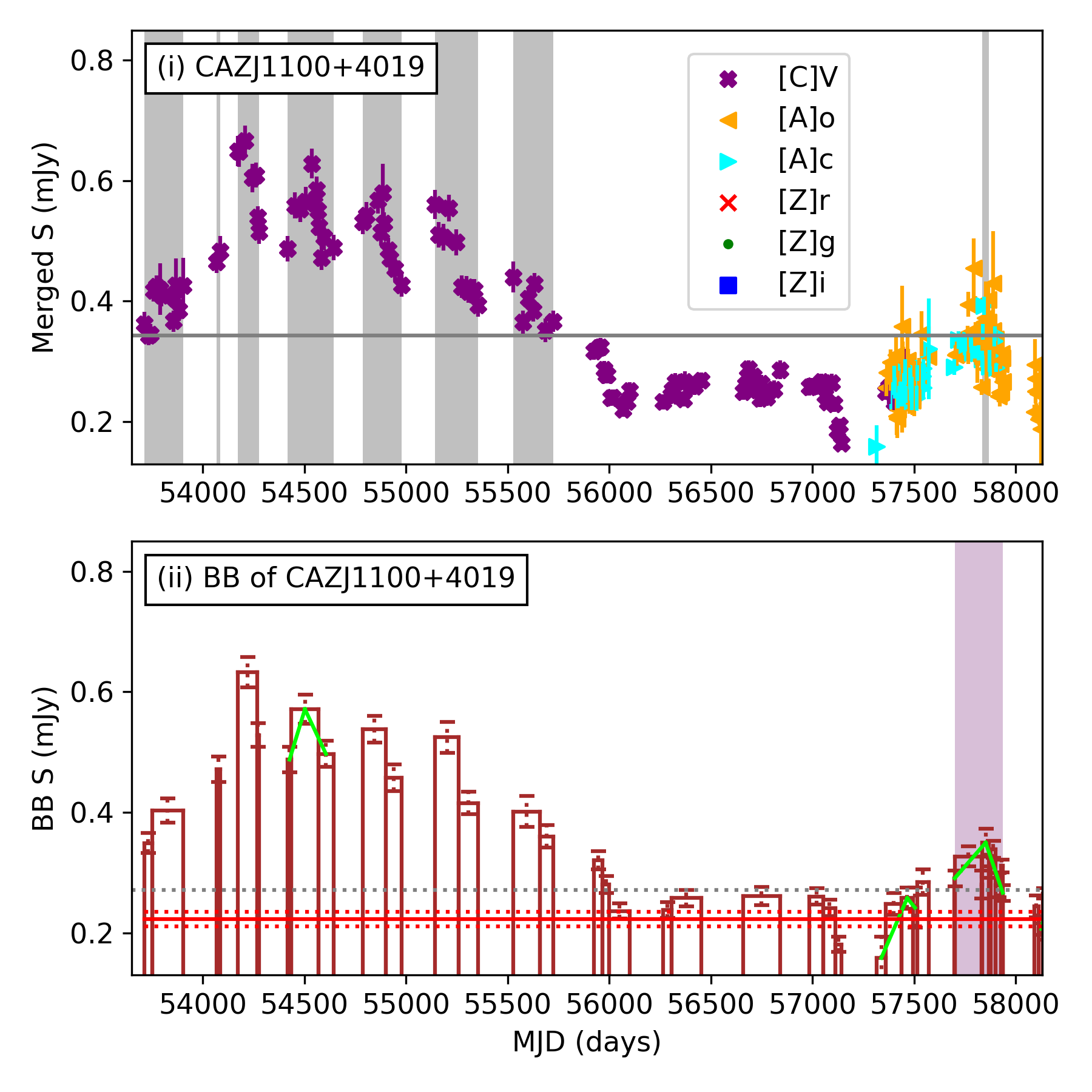}
    \caption{Difference between BB95 (gray vertical areas in panel i), BBHOP flares (green lines in panel ii), and prominent BBHOP flares (purple area along with green lines in panel ii). Panel (i): Merged CAZ light curve. Panel (ii): BBs of the merged CAZ light curve for a zoomed-in portion of CAZJ1100+4019. The horizontal solid gray line represents $S_{95\%}$, the horizontal dotted gray line represents $S_{75\%}$, and the horizontal solid red line shows $S_{25\%}$ (with its errors shown as dotted red lines). For more plot details, see the description of Fig. \ref{fig_eg_LCs_eg_flares}. The BBHOP algorithm fails to find flares in low cadence periods (e.g., CRTS-epoch) due to frequent observational gaps. Notably, unlike the BBHOP algorithm, the performance of the BB95 algorithm on periods with data is not affected by the presence of neighboring observational gaps.}
    \label{fig_eg_LCs_eg_flares_CAZJ1100}
\end{figure}

\begin{figure}
    \centering
    \includegraphics[width=0.49\textwidth]{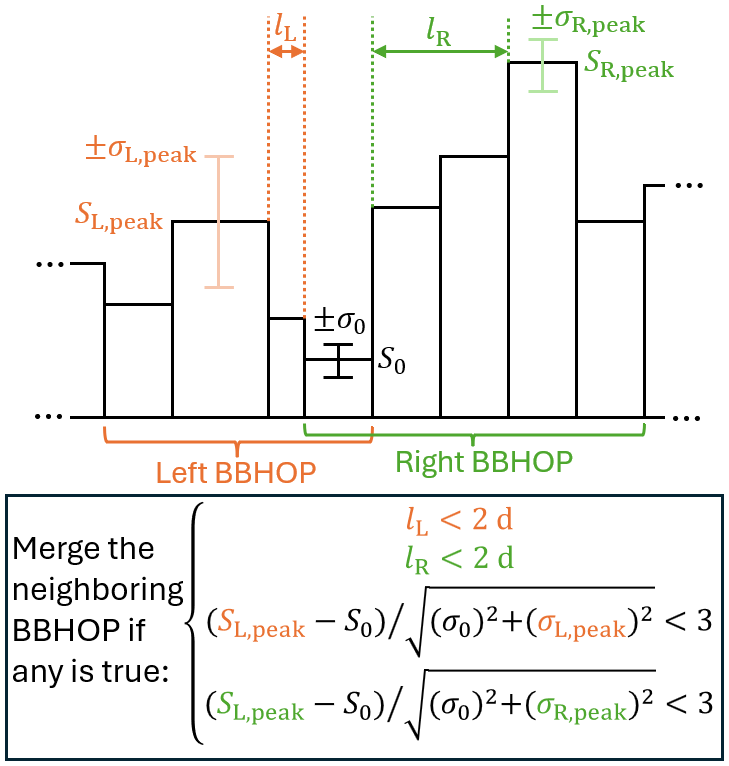}
    \caption{Merging two neighboring BBHOP flares. If any of the four conditions is true, the two neighboring BBHOP flares are merged into a single BBHOP flare. The start BB, end BB, and peak BB of the merged flare are the start BB of the left BBHOP, end BB of the right BBHOP, and the larger of the two peak BBs, respectively.}
    \label{fig_sketch_flare_merging}
\end{figure}

\begin{figure}
    \centering
    \includegraphics[width=0.49\textwidth]{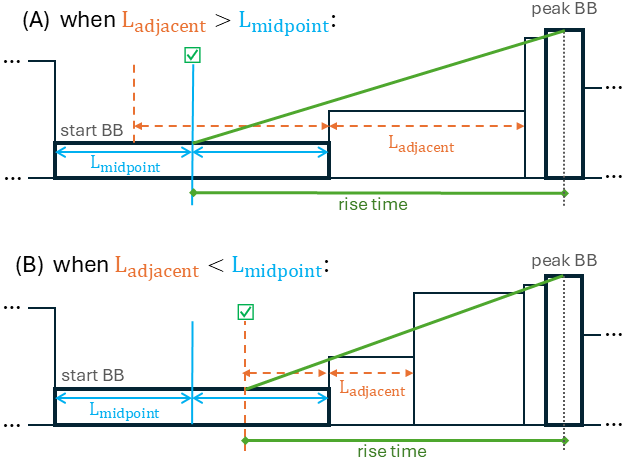}
    \caption{Scenarios used to determine the start time of a BBHOP flare. In scenario (A), the midpoint of the start BB is chosen as the start, as it minimizes the flare duration. Conversely, in scenario (B), the duration of the adjacent BB is used to obtain the start, as flipping it minimizes the flare duration. A mirrored equivalent of this principle is used for determining the fall time of a flare.}
    \label{fig_sketch_flare_start}
\end{figure}

\begin{figure*}
    \centering
    \includegraphics[width=17cm]{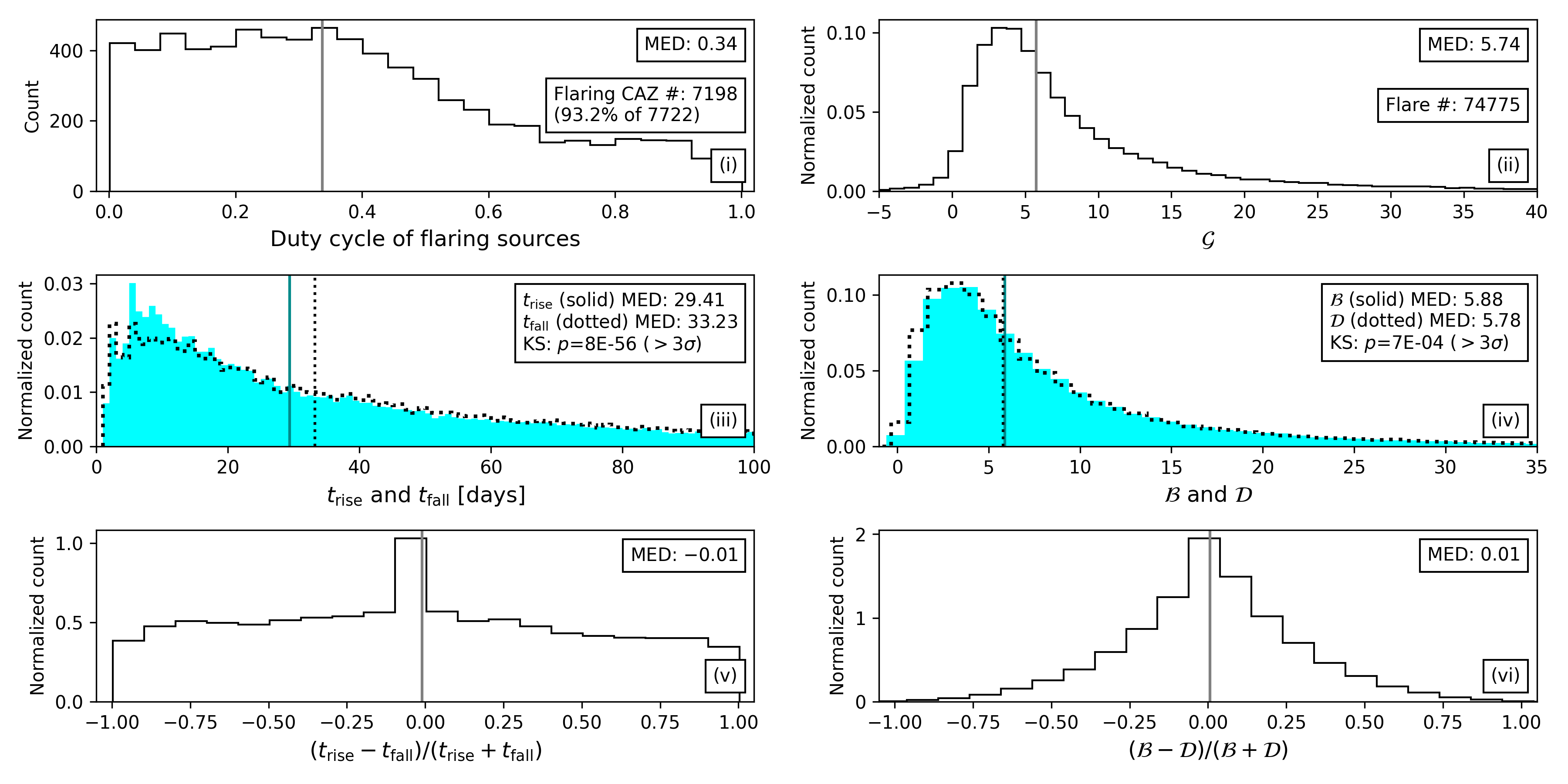}
    \caption{Flare characteristics of all 74775 BBHOP flares. The panels are as follows:(i) Distribution of the duty cycle of flaring sources. (ii) Global flare amplitudes ($\mathcal{G}$). (iii) Rise and fall times of flares ($t_\mathrm{rise}$ and $t_\mathrm{fall}$). (iv) Local brightening and decaying flare amplitudes ($\mathcal{B}$ and $\mathcal{D}$). (v) Temporal asymmetry of flares. (vi) Amplitude asymmetry of flares. In panel (iii) we perform a KS test on the distributions of $t_\mathrm{rise}$ and $t_\mathrm{fall}$ and do the same on the distributions of $\mathcal{B}$ and $\mathcal{D}$ in panel (iv). The term ``MED'' refers to the median of each distribution and is visualized using a vertical line.}
    \label{fig_flares_all_CAZ_ALL}
\end{figure*}

\begin{figure*}
    \centering
    \includegraphics[width=17cm]{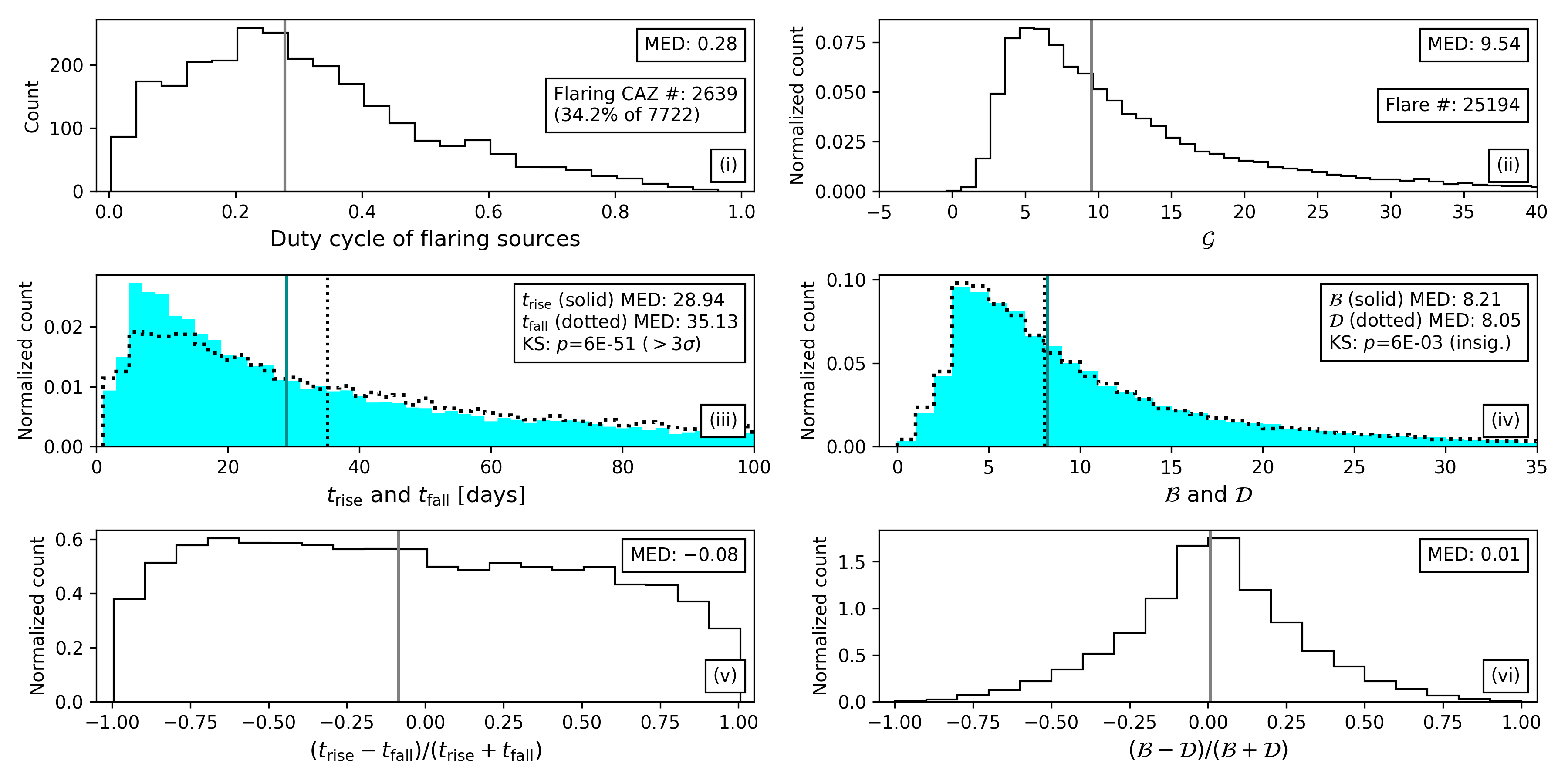}
    \caption{Flare characteristics of 25194 prominent BBHOP flares. For plot details, see the description of Fig. \ref{fig_flares_all_CAZ_ALL}.}
    \label{fig_flares_all_CAZ_TOP}
\end{figure*}

While the BBHOP algorithm identifies the majority of flaring periods, its reliability is limited by the cadence of the data. In epochs with low data count and frequent observational gaps (e.g., the CRTS-epoch of the merged CAZ light curves), the BBHOP algorithm cannot easily find valley-hill-valley BBs and thus BBHOP flares. This is not a limitation for the BB95 algorithm (see Sect. \ref{sec_identifying_critically_bright_BBs}), thus offering an advantage over the BBHOP algorithm in such low cadence periods (e.g., Fig. \ref{fig_eg_LCs_eg_flares_CAZJ1100}), even though BBHOP flares do in general offer a more complete picture of the flaring activity.

Despite increasing the \texttt{ncp\_prior}$_\mathrm{ideal}$ (see Appendix \ref{appendix_BB_calib}), the BB algorithm remains rather sensitive to fast variations in the merged CAZ light curves, especially in the more noisy ATLAS+ZTF-epoch. Fast variations can prematurely break valley-hill-valley BB patterns. This becomes detrimental when the BBHOP algorithm is tracing major outburst activity over longer rise and fall times. Moreover, blazars commonly show small flux variations during major outbursts, which also break the BBHOP continuity when tracing the underlying flare. To ensure that our BBHOP flares generally correspond to major flaring periods, we merge two BBHOP flares if: (1) the time difference from the peak BB to end BB of the first one is faster than 2~d; (2) the flux density decay of the first one is less significant than 3$\sigma$; (3) the time difference from the start BB to peak BB of the second one is faster than 2~d; or (4) the flux density rise of the second one is less significant than 3$\sigma$. These criteria are visualized in Fig. \ref{fig_sketch_flare_merging}. When two BBHOP flares are merged, they form a single BBHOP flare: the start BB of the first one becomes the start BB of the merged flare, the end BB of the second one the end BB of the merged flare, and the larger of the two peak BBs is chosen as the peak BB of the merged flare. We iteratively merge all neighboring BBHOP flares (including newly merged ones with their potential neighbor) until neighboring BBHOP flares cannot be merged. Henceforth, the term ``BBHOP flare'' also refers to merged BBHOP flares.

The peak time of a BBHOP flare is defined as the middle of the peak BB. The start and end times of a flare are chosen using one of two methods. In the first method, the flare starts at the midpoint of the start BB. In the second method, we identify the BB adjacent to the start BB and use its duration to trace back in time from the start BB to obtain a start time for the flare (i.e., the adjacent block is ``flipped''; e.g., \citealt{Wagner2022_BBHOP_flips}). Between these two start times, we choose the one that minimizes the total duration of a flare. The end time of a flare is determined following the same logic, but mirrored. This minimization ensures that the flares do not have unnaturally long start or end times when they are preceded or succeeded by an unusually long quiescent period, respectively. These two scenarios are respectively sketched in panels (A) and (B) of Fig. \ref{fig_sketch_flare_start}.

We define ``brightening'' ($\mathcal{B}$) as the significance of flux density change\footnote{Significance of change from $S_1 \pm \sigma_1$ to $S_2 \pm \sigma_2$ is $\frac{S_2-S_1}{\sqrt{\sigma_1^2 + \sigma_2^2}}$.} of the start BB to the peak, and ``decaying'' ($\mathcal{D}$) as that of the end BB to the peak. In addition to these significances of local flux density change, we define $\mathcal{G}$ as the significance of flux density change from a global quiescent-level to the peak. The global quiescent-level is defined as the median of the lowest 50\% of the merged flux densities (i.e., $S_{25\%}$). In this work, we refer to $\mathcal{B}$, $\mathcal{D}$, and $\mathcal{G}$ as flare ``amplitudes'', which should not be confused with the convention where the flare amplitude is defined as its flux density change. Nonetheless, $\mathcal{B}$, $\mathcal{D}$, and $\mathcal{G}$ are directly proportional to flux density change, but additionally depend on the flux density errors.

We defined $t_\mathrm{rise}$ as the time taken between the start and peak of a flare (see Fig. \ref{fig_sketch_flare_start}), and $t_\mathrm{fall}$ between its peak and end. We investigate the asymmetry in the rise and fall times using $(t_\mathrm{rise}-t_\mathrm{fall})/(t_\mathrm{rise}+t_\mathrm{fall})$. Likewise, we inspect for asymmetry in the local brightening and decaying amplitudes using $(\mathcal{B}-\mathcal{D})/(\mathcal{B}+\mathcal{D})$. We also define the ``duty cycle'' of a source as the ratio of the total flaring time to the total duration of all BBs in a light curve. The duration of gaps in the BB light curve is omitted from the duty cycle calculation.

We looked for BBHOP flares in 7722 (97.5\% of 7918 AGN) merged light curves with $\ge$10 data points and $\ge$30~d duration. We find 74775 BBHOP flares across 7198 (93.2\% of 7722) light curves. In Fig. \ref{fig_flares_all_CAZ_ALL} we show the global distributions of the duty cycle, $\mathcal{G}$, $t_\mathrm{rise}$ and $t_\mathrm{fall}$, $\mathcal{B}$ and $\mathcal{D}$, as well as the temporal and amplitude asymmetry parameters.

Despite calibrating the Bayesian prior of the BB algorithm using WDs, we find that many BBHOP flares arise due to noise-driven scatter (e.g., see Fig. \ref{fig_eg_LCs_eg_flares} where we show a zoomed-in version of Fig. \ref{fig_eg_LCs}). This is because: (1) the automatic forced-photometry and reduction algorithms of the all-sky surveys (especially those of ATLAS) cannot always reliably quantify the flux density and flux density error of blazars, particularly those which are host-galaxy-dominated,\footnote{Correcting for the effect of host-galaxy contamination on a blazar light curve requires meticulous modeling of the brightness of the host galaxy itself (e.g., \citealt{Nilsson2007_host_galaxy}). This necessitates carefully centering an appropriately sized photometry aperture on the source, a process which is also sensitive to atmospheric conditions (i.e., the ``seeing''). Additionally, the aperture size depends on the intrinsic brightness and angular size of the host galaxy, which should be determined on a source-by-source basis. Such careful host-galaxy modeling is not possible when using forced-photometry data and thus lies beyond the scope of this work.} resulting in extreme outliers; (2) the merged light curves combine data from three such surveys, with differing observational properties (e.g., limiting magnitudes, error estimations); and (3) blazars often exhibit complex color variations and intrinsic properties across different filters which we could not account for when merging them. Therefore, when investigating the flaring properties of the two main blazar populations (FSRQs and BLLs) with respect to some physical properties (see Sect. \ref{sec_results_flares}), we mainly focus on the most prominent BBHOP flares.

We selected a prominent flare by requiring that (1) its light curve is variable (2798 of 7722 are; see Sect. \ref{sec_variability}), (2) its peak BB is brighter than the 75$^\mathrm{th}$ percentile flux density (i.e., $S_{75\%}$), (3) one of its local amplitudes (i.e., $\mathcal{B}$ or $\mathcal{D}$) is larger than 3, and (4) it is not a 3-block flare with only one data point in its peak BB. The first condition excludes 34780 (46.5\% of 74775) flares, the second another 11333 (15.2\%), the third another 2145 (2.9\%), and the fourth another 1323 (1.8\%). Thus, in total there are 25194 (33.7\%) prominent BBHOP flares, whose global distributions are shown in Fig. \ref{fig_flares_all_CAZ_TOP}. In Sect. \ref{sec_results_flares} we discuss and interpret Figs. \ref{fig_flares_all_CAZ_ALL} and \ref{fig_flares_all_CAZ_TOP}. Examples of BBHOP and prominent BBHOP flares are shown in Figs. \ref{fig_eg_LCs_eg_flares} and \ref{fig_eg_LCs_eg_flares_CAZJ1100}.

\section{Results and discussion} \label{sec_results}
Here we describe and discuss the results of our variability and flare analyses of the CAZ light curves. In some parts, we only focus on the two main blazar subpopulations, FSRQs and BLLs (see Sect. \ref{sec_intro}), of which we have 2495 and 2726 sources, respectively (see Sect. \ref{sec_compiling_caz_sources}). In Sect. \ref{sec_results_variability} their overall variability via $F_\mathrm{var}$ is investigated and interpreted, in Sect. \ref{sec_results_critically_bright_BB} their periods of enhanced emission via BB95, and in Sect. \ref{sec_results_flares} their flaring behavior via BBHOP.

\subsection{Variability of CAZ light curves} \label{sec_results_variability}
We measure the overall variability of a light curve using $F_\mathrm{var}$. As mentioned in Sect. \ref{sec_variability}, the inclusion of the ATLAS data in the merged CAZ light curves introduces non-intrinsic variability. Thus, we investigate the behavior of blazar variability with respect to synchrotron peak frequency and radio variability Doppler factor only using survey-specific (CRTS- and ZTF-only) merged light curves. Since these distributions for CRTS- and ZTF-only light curves are comparable and lead to identical conclusions, for brevity, we just present the results for ZTF-only light curves.

In addition to $F_\mathrm{var}$, we also investigate the behavior of median flux density against the above physical parameters as well as redshift. A caveat associated with this analysis is the shifted nature of the filters (see Sect. \ref{sec_characterizing_variability}). Even when we only focus on ZTF-only merged light curves, some shifting issues are present since [Z]r, [Z]g, and [Z]i are shifted onto each other (in order of decreasing priority; see Sect. \ref{sec_merge_all_filters}).

\begin{figure*}[h!]
    \centering
    \includegraphics[width=18cm]{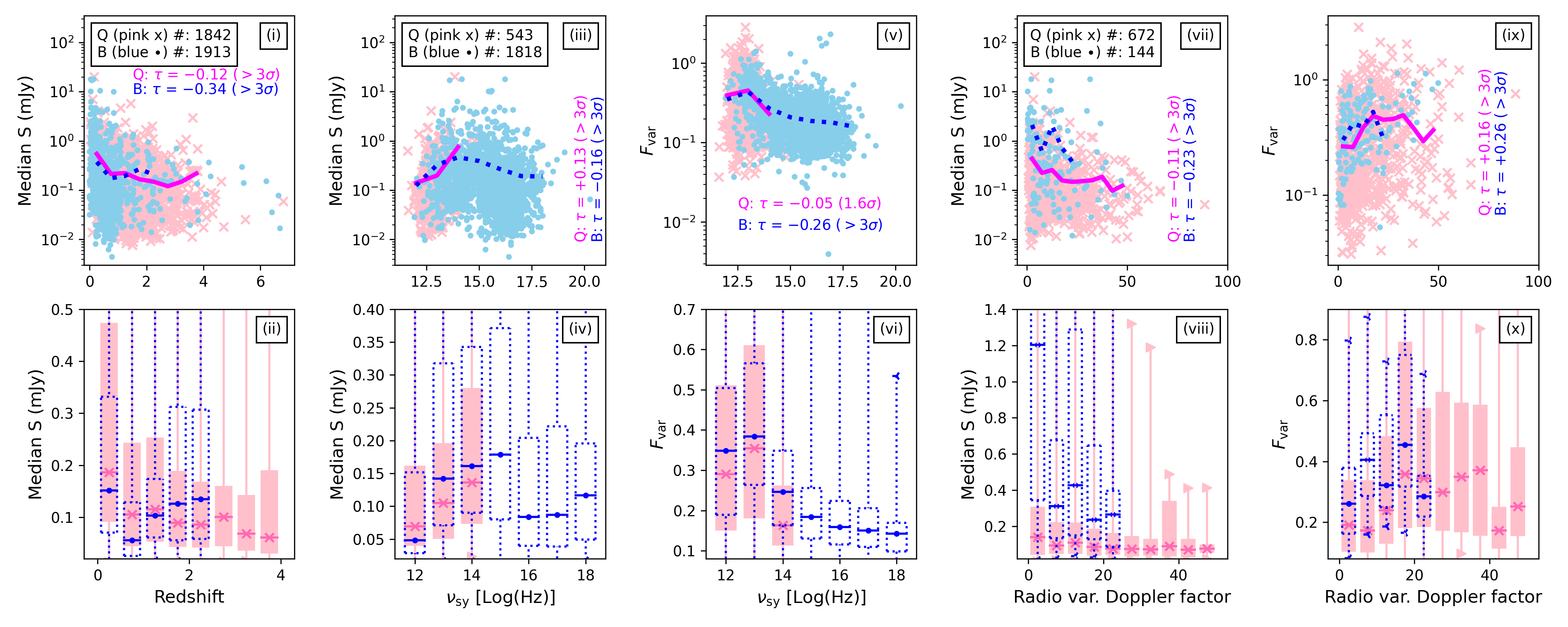}
    \caption{Distribution of median flux density against redshift (i and ii), median flux density and $F_\mathrm{var}$ against synchrotron peak frequency (iii--vi), and median flux density and $F_\mathrm{var}$ against radio variability Doppler factor (vii--x) in ZTF-only merged light curves of the sources in the CAZ catalog. Pink crosses and solid markings refer to FSRQs (Q); blue dots and dotted markings refer to BLLs (B). The top panels are scatter plots (with y-axis in logarithmic scales), where the binned running average is represented using lines (solid for FSRQs and dotted for BLLs). Kendall's $\tau$ correlation coefficient and significance are reported for FSRQs and BLLs. In the bottom panels, we show the binned box plots of each scatter plot. The averages and box plots are both only shown for bins with at least ten data points. In the top panels we show all data points, while in the bottom panels we zoom in on the boxes.}
    \label{fig_flux_Fvar_ZTF}
\end{figure*}

\subsubsection{Flux density against redshift} \label{sec_results_variability_vs_redshift}
In Fig. \ref{fig_flux_Fvar_ZTF} (i and ii), and all such plots in this section, we show the distribution of median flux density of the two main blazar populations (FSRQs and BLLs) against redshift ($z$). In panel (i), we calculate Kendall's $\tau$ non-parametric correlation coefficient for the FSRQ and BLL scatter plots separately. $\tau$ ranges from $+1$ to $0$ then to $-1$, indicating a perfect positive correlation, no correlation, and a perfect negative correlation, respectively. The significance of $\tau$ is given in parentheses in terms of Gaussian $\sigma$. In panel (ii), we draw box plots and zoom in. A box extends from the 25$^\mathrm{th}$ percentile to the 75$^\mathrm{th}$ percentile of the binned data, with the median (50$^\mathrm{th}$ percentile) being marked in between them. The lines below and above the box extend until the minimum and maximum values, respectively. In the following discussions we generally focus on the behavior of the boxes rather than the extensions. We note that the box plots as well as the running averages in the scatter plots are shown only when there are at least ten data points in each bin.

As seen from Fig. \ref{fig_flux_Fvar_ZTF} (i and ii), we find an overall decreasing trend in blazar brightness as redshift increases as evident from the $>$3$\sigma$ mildly negative $\tau$ correlations of $-0.12$ and $-0.34$ for FSRQs and BLLs, respectively. Interestingly though, this decrease in brightness flattens after $z \sim 0.5$, similar to \cite{Abrahamyan2019_mag_vs_z}. This flattening is also seen when only the more reliable spectroscopic redshifts are used. This apparent disagreement with the inverse-square law dimming is not unexpected as many factors affect blazar luminosity, namely: an observation bias toward greater relativistic boosting at larger redshifts (e.g., \citealt{Lister1997_blz_cosmology, Hovatta2009_DopF_vs_z}; also see Sect. \ref{sec_results_variability_vs_DopF}), and an increase in AGN density and activity at $z \sim 2$ (e.g., \citealt{Hopkins2007_peak_AGN_activity}). The trend is even more complicated in the optical band. Firstly, at lower redshifts the effect of host-galaxy contribution is more prominent (especially in BLLs; e.g., \citealt{Kotilainen1998_host_brightness_vs_z}). We tested this using the methodology of \cite{Tachibana2018_PanSTARRS_host_dominated_analysis}, confirming that a larger fraction of our sources at $z \lesssim 0.5$ are galaxy-like. This is partly why we have brighter sources at $z \lesssim 0.5$. Secondly, the behavior of optical brightness and variability of a blazar against redshift depends strongly on its SED class and shape (i.e., spectral index), because at higher redshifts the observed photons originate from an increasingly higher frequency part of the SED (see Sect. \ref{sec_results_variability_vs_nu}). Accounting for this effect requires K-correction (e.g., \citealt{Richards2006_K_correction}) which cannot be coherently performed for such a large number of multi-filter light curves and is thus beyond the scope of this work. Therefore, in this paper we do not investigate variability and flare characteristics of the CAZ light curves against redshift.

\subsubsection{Flux density and variability against synchrotron peak frequency} \label{sec_results_variability_vs_nu}
Figure \ref{fig_flux_Fvar_ZTF} (iii--vi) shows the median flux density and $F_\mathrm{var}$ of ZTF-only merged light curves against synchrotron peak frequency. As expected, higher synchrotron peak frequencies (ISP and HSP regimes) are only occupied by BLLs, since FSRQs are generally LSPs whereas BLLs can range from LSPs to EHSPs (Sect. \ref{sec_intro}).

Emissions originating from closer to the synchrotron peak are brighter and more variable, due to the electrons having the highest energies, cooling the fastest, and being closest to the acceleration zones. This leads to a universal decrease in brightness and variability as lower and higher frequencies than $\nu_\mathrm{sy}$ are probed (e.g., \citealt{Ikejiri2011_var_v_nu, Ackermann2011_fermi, Hovatta2014_opt_vs_gamma_variability, Richards2014_radio_var, Balokovic2016_higher_energy_electron_more_variable, Itoh2016_var_v_nu, Safna2020_var_vs_nu, OteroSantos2023_var_and_PD}). This is what we see in Fig. \ref{fig_flux_Fvar_ZTF}, where both FSRQs and BLLs peak in flux density at $\nu_\mathrm{sy} \sim 10^{14}$~Hz, and in $F_\mathrm{var}$ at $\nu_\mathrm{sy} \sim 10^{13}$~Hz, both of which are generally comparable to the optical band frequency of $\sim$$10^{14}$~Hz. We note that these trends are  supported by the significant ($>$3$\sigma$) $\tau$ correlations present for FSRQs and BLLs in their respective $\nu_\mathrm{sy}$ range. Interestingly, FSRQs and BLLs generally behave the same at the same $\nu_\mathrm{sy}$ range, in agreement with \cite{Hovatta2014_opt_vs_gamma_variability}.

\subsubsection{Flux density and variability against radio variability Doppler factor} \label{sec_results_variability_vs_DopF}
In Fig. \ref{fig_flux_Fvar_ZTF} (vii--x) we explore how median flux density and $F_\mathrm{var}$ behave in terms of radio variability Doppler factor, which is calculated using flare amplitudes in the radio band (e.g., \citealt{liodakis2018_VarDopplerFac}; also see Sect. \ref{sec_compiling_caz_sources}). Doppler factor generally quantifies the effect of relativistic boosting and is obtained using $\delta = [ \, \Gamma (1-\beta \cos(\theta))]^{-1}$ where $\Gamma = (1-\beta^2)^{-1/2}$ is the Lorentz factor, $\beta$ the bulk plasma speed in $c$, and $\theta$ the viewing angle (e.g., \citealt{blandford2019}). Therefore, as the bulk plasma speed increases and as the viewing angle decreases, Doppler factor increases. An increased Doppler factor results in greater relativistic boosting of the observed luminosities and variability amplitudes (e.g., \citealt{Blandford1979}). This leads to the observational bias of having a greater fraction of high Doppler factor blazars at larger redshifts.

Since a larger radio variability Doppler factor implies an enhancement of relativistic boosting due to physical (i.e., increased bulk Lorentz factor) and geometrical (i.e., smaller viewing angle) properties of the jet, emissions in other bands should also experience similar enhancements. We see a hint of this with the $>$3$\sigma$ mildly positive $\tau$ correlations of $+0.16$ and $+0.26$ for FSRQs and BLLs, respectively, suggesting that $F_\mathrm{var}$ increases as radio variability Doppler factor increases. Nonetheless, the increasing trend appears to flatten at Doppler factor $\gtrsim20$, which may be due to lower source counts. Further evidence of this trend is provided in Sect. \ref{sec_results_flares_v_DopF} using BBHOP flares.

Surprisingly, flux density shows a general drop as the radio variability Doppler factor increases ($>$3$\sigma$ mildly negative $\tau$ correlations of $-0.11$ and $-0.23$ for FSRQs and BLLs, respectively). This is likely due to increased host-galaxy contamination for sources with smaller Doppler factors as they are most often found at low redshifts. This effect is especially noticeable for BLLs. We note that this could also be a reason why $F_\mathrm{var}$ is lower at smaller Doppler factors, since host-galaxy dilution can suppress variability. Another reason for the decreasing trend of median flux density with Doppler factor could be due to the aforementioned observational bias. In other words, a larger fraction of high Doppler factor sources are expected to be at higher redshifts and thus dimmer in flux density.

\subsection{Periods of enhanced emission in CAZ light curves} \label{sec_results_critically_bright_BB}
\begin{figure}
    \centering
    \includegraphics[width=0.49\textwidth]{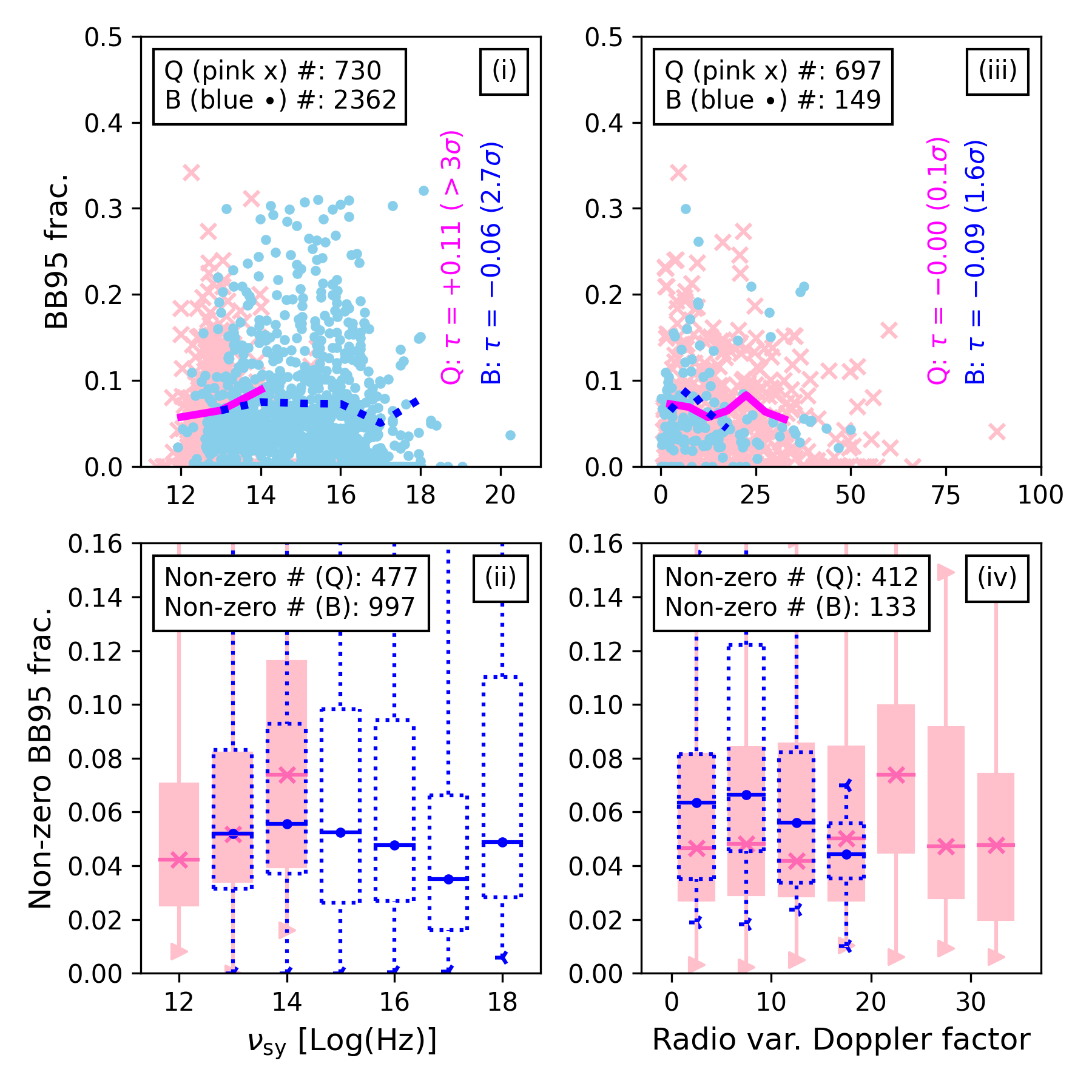}
    \caption{Distribution of BB95 (with respect to all BB) fraction against synchrotron peak frequency (i and ii) and radio variability Doppler factor (iii and iv). Pink crosses and solid markings refer to FSRQs (Q); blue dots and dotted markings refer to BLLs (B). Kendall's $\tau$ correlation coefficient and significance are reported for FSRQs and BLLs. If a source has a BB95 fraction of zero, it is plotted in the top panels but not in the bottom ones. The running averages in the top row (shown as solid lines for FSRQs and dotted lines for BLLs) and the box plots in bottom row are both only calculated for at least ten nonzero values within each bin. In the top panels we show all data points, while in the bottom panels we zoom in on the boxes.}
    \label{fig_critically_bright_BB_frac_TOP}
\end{figure}

\begin{figure*}[h!]
    \centering
    \includegraphics[width=18cm]{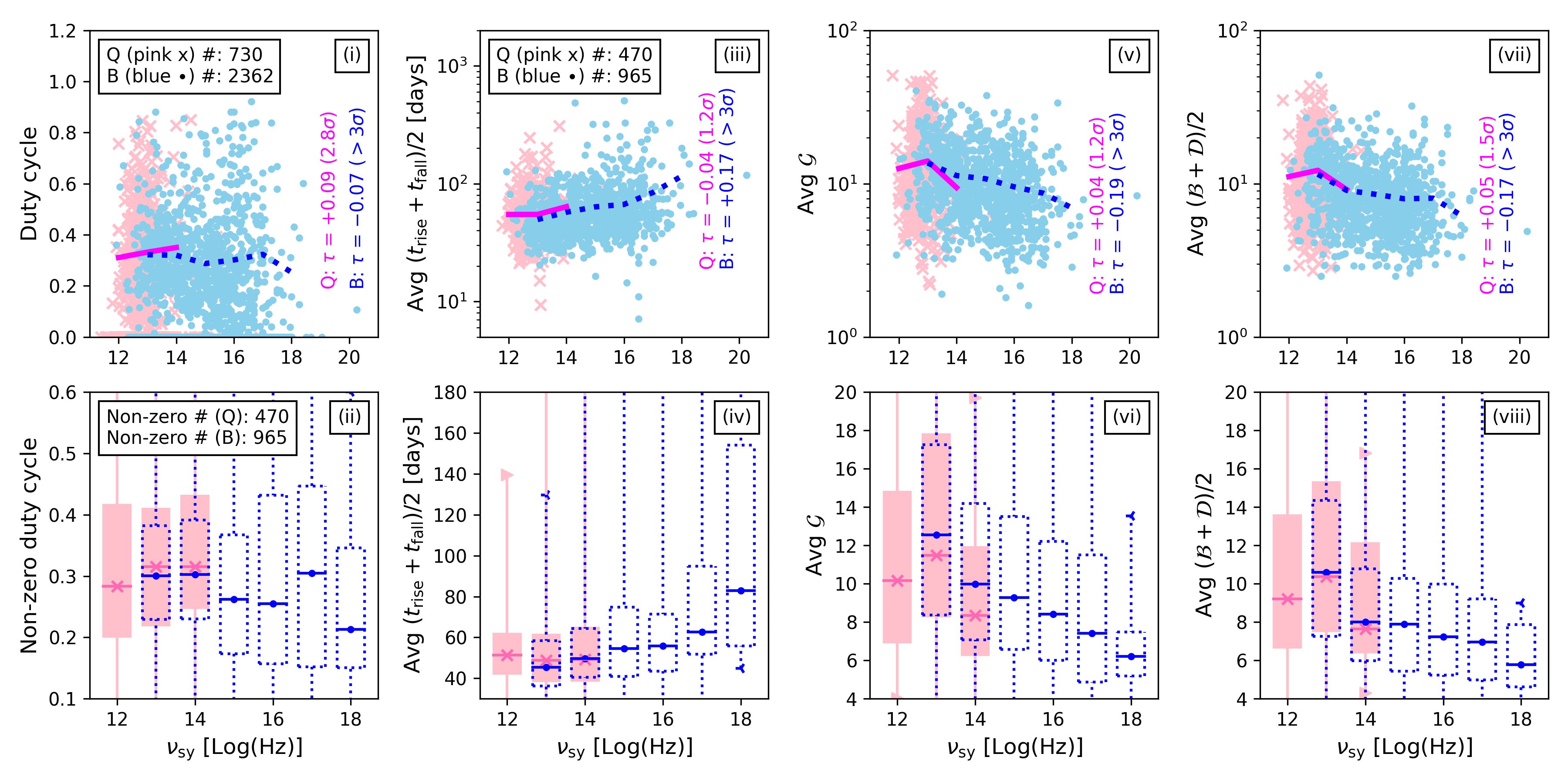}
    \caption{Distribution of source duty cycle (i and ii) as well as source-averaged average rise and fall times (iii and iv), global amplitude (v and vi), and average local brightening and decaying amplitudes (vii and viii) of prominent BBHOP flares against synchrotron peak frequency. Pink crosses and solid markings refer to FSRQs (Q); blue dots and dotted markings refer to BLLs (B). Kendall's $\tau$ correlation coefficient and significance are reported for FSRQs and BLLs. In panel (i) we show zero duty cycle data, whereas in the other panels we do not. The running averages in the top row (shown as solid lines for FSRQs and dotted lines for BLLs) and the box plots in bottom row are both only calculated for at least ten nonzero values within each bin. In the top panels we show all data points, while in the bottom panels we zoom in on the boxes.}
    \label{fig_LogLE_v_SourceAVG_flare_TOP}
\end{figure*}

We plot the distribution of BB95 to all BB fraction against synchrotron peak frequency and radio variability Doppler factor in Fig. \ref{fig_critically_bright_BB_frac_TOP}. We note that these BB95 are limited to those from variable CAZ light curves (see Sect. \ref{sec_variability}). The BB95 fraction of FSRQs shows a $>$3$\sigma$ weakly positive trend ($\tau=+0.11$) and peaks at $\nu_\mathrm{sy} \sim 10^{14}$. For BLLs, the median of the BB95 fraction is also greatest at $\nu_\mathrm{sy} \sim 10^{14}$. However, its trend is mostly flat against $\nu_\mathrm{sy}$ ($\tau=-0.06$ at a significance of 2.7$\sigma$). As such, we cannot significantly determine whether the fraction of BB95 peaks at $\nu_\mathrm{sy} \sim 10^{14}$. If it did, it would suggest that blazars whose synchrotron emission peaks at the optical energy exhibit more frequent periods of enhanced emission when probed at the optical band, which would fall inline with the reasoning of Sect. \ref{sec_results_variability_vs_nu}. Interestingly, a similar behavior is seen with BBHOP duty cycles against $\nu_\mathrm{sy}$ (see Fig. \ref{fig_LogLE_v_SourceAVG_flare_TOP} panels i and ii). Nevertheless, we reiterate that this trend cannot be confirmed. 

As seen in Fig. \ref{fig_critically_bright_BB_frac_TOP} (iii--iv), BB95 fraction does not seem to depend on radio variability Doppler factor. This is expected as an increased Doppler factor does not imply an increased frequency of flaring in blazars. We explore this further using the duty cycle of BBHOP flares in Sect. \ref{sec_results_flares_global_duty_cycle}.

\subsection{Flaring behavior in CAZ light curves} \label{sec_results_flares}
We identify flares in merged CAZ light curves by employing the BBHOP algorithm (see Sect. \ref{sec_identifying_flares}). In total we find $\sim$75k BBHOP flares across 7198 light curves. The global distributions of these flares are given in Fig. \ref{fig_flares_all_CAZ_ALL}. As discussed in Sect. \ref{sec_identifying_flares}, many of these are noise-like and cannot always be confidently identified as intrinsic flares. As a result, we only focus on $\sim$25k prominent flares across 2639 light curves to investigate the physical nature of blazar flares. The global distributions of the prominent BBHOP flares are shown in Fig. \ref{fig_flares_all_CAZ_TOP}.

\begin{figure*}[h!]
    \centering
    \includegraphics[width=18cm]{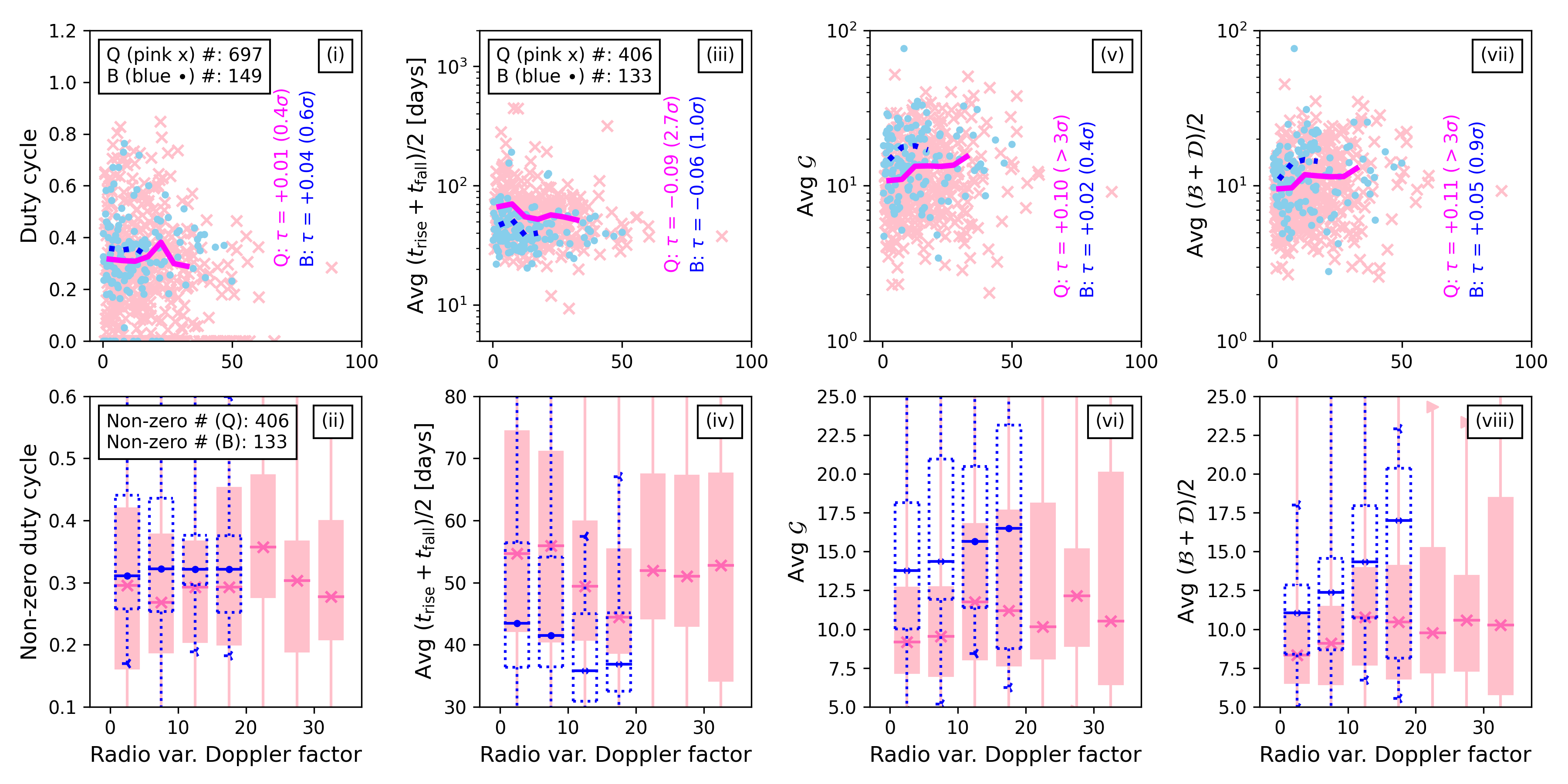}
    \caption{Similar to Fig. \ref{fig_LogLE_v_SourceAVG_flare_TOP} but for the radio variability Doppler factor. We show similar plots using redshift-corrected parameters in Appendix \ref{appendix_z_corr_Doppler_factor}.}
    \label{fig_DopF_v_SourceAVG_flare_TOP}
\end{figure*}

\subsubsection{Global BBHOP flare characteristics} \label{sec_results_flares_global}
Here we discuss the distributions of the duty cycle, amplitude, and rise and fall times of BBHOP flares, which are found in Figs. \ref{fig_flares_all_CAZ_ALL} and \ref{fig_flares_all_CAZ_TOP}.

\subsubsubsection{Duty cycles of BBHOP flares} \label{sec_results_flares_global_duty_cycle}
As evident from Fig. \ref{fig_flares_all_CAZ_ALL}, the majority (93.2\%) of CAZ light curves have at least one BBHOP flare. However, only 34.2\% have at least one prominent flare (see Fig. \ref{fig_flares_all_CAZ_TOP}). When limited to prominent flares, the duty cycle (fraction of flaring time to the total BB time) is generally small ($\lesssim$0.3), but can be as large as 0.9. Such a wide range of duty cycles is in agreement with \cite{Bauer2009_PTF}, who studied optical variability of 362 blazars.

We note that large BBHOP duty cycles are possible because the entire duration of a BBHOP flare, including the low flux rising or falling portions, is used when calculating the total flaring time. Thus, light curves that exhibit back-to-back major flares with rather extended rise and fall times are likely to have very large BBHOP duty cycles. For the fraction of time CAZ light curves spend at high flux density, we refer the reader to BB95 fraction (see Sect. \ref{sec_results_critically_bright_BB}). Nevertheless, we caution that some large duty cycles arise due to non-intrinsic flux density variations in problematic light curves. Despite our light curve cleaning efforts and stringent selection criteria for the prominent BBHOP flares, not all of them are intrinsic and their overall duration can become greatly exaggerated (especially as a result of BBHOP merging; see Sect. \ref{sec_identifying_flares}).

\subsubsubsection{Amplitudes of BBHOP flares} \label{sec_results_flares_global_amplitudes}
Global amplitude ($\mathcal{G}$, measured as the flux density significance change from the peak BB to a global quiescence level in Sect. \ref{sec_identifying_flares}) peaks at $\sim$3 for all flares (Fig. \ref{fig_flares_all_CAZ_ALL} plot ii) and $\sim$5 for prominent flares (Fig. \ref{fig_flares_all_CAZ_TOP} plot ii). Meanwhile, the local amplitudes (brightening $\mathcal{B}$ and decaying $\mathcal{D}$, measured as the flux density significance change from the peak BB to the starting and ending BBs, respectively) peak at $\sim$3.1 for all flares (Fig. \ref{fig_flares_all_CAZ_ALL} plot iv) and at $\sim$3.6 for prominent ones (Fig. \ref{fig_flares_all_CAZ_TOP} plot iv). These trends are expected based on the selection criteria of the prominent flares.

Regarding amplitude asymmetry, as seen in plot (vi) of Figs. \ref{fig_flares_all_CAZ_ALL} and \ref{fig_flares_all_CAZ_TOP} via the $(\mathcal{B}-\mathcal{D})/(\mathcal{B}+\mathcal{D})$ distributions and the two-sample Kolmogorov-Smirnov (KS) tests, $\mathcal{B}$ and $\mathcal{D}$ are generally symmetrical (the $>$3$\sigma$ KS test becomes insignificant when going from all to prominent flares). This intuitive result indicates that flaring states in blazars are transient and all blazars eventually return to a quiescent state.

\subsubsubsection{Timescales of BBHOP flares} \label{sec_results_flares_global_timescales}
The rise and fall times ($t_\mathrm{rise}$ and $t_\mathrm{fall}$, respectively) of BBHOP flares peak at $\sim$7~d. These timescales are generally comparable the $\sim$1--8~d characteristic timescales found by \cite{Dingler2024_optical_characteristics_timescales} for 67 blazars. The rise and fall times exhibit a substantial spread, with their 25$^\mathrm{th}$ and 75$^\mathrm{th}$ percentiles being $\sim$13--63~d and $\sim$16--71~d, respectively. This translates to typical BBHOP flare durations of $\sim$30--130~d. The large spread is compatible with the characteristic timescales found by \cite{Xiong2025_ZTF_var_timescales} for 1684 blazar ZTF light curves. Interestingly, BBHOP flares are clearly asymmetric in time (KS p-values of $<$$10^{-50}$, i.e., KS significance of $\gg$3$\sigma$) with a faster rise than fall. For example, the median rise time is $\sim$29~d while the median fall time $\sim$35~d. The distribution of $(t_\mathrm{rise}-t_\mathrm{fall})/(t_\mathrm{rise}+t_\mathrm{fall})$ is skewed toward the negative values with a median of $-0.08$.

Blazars are known to show both temporally symmetrical (e.g., \citealt{Chatterjee2012_flare_symmetry, Blinov2015_symmetric_optical_flares, Cerruti2025_Fermi_flare_temporal_symmetry}) and asymmetrical flares (e.g., \citealt{Nalewajko2013_gamma_asymmetry, Roy2019_flare_symmetry, Valtaoja1999_faster_rise, Nieppola2009_faster_rise}) in multiple bands. Temporal symmetry generally arises when light crossing time dominates the shape of the flares, while asymmetry from rapid acceleration or cooling of the emitting particles. Both symmetrical and asymmetrical flares can be explained by blazar acceleration models (with shock recollimation models, e.g., \citealt{Fromm2016_shock_model, FichetdeClairfontaine2021_shock_model}; magnetic reconnection, e.g., \citealt{Cerutti2012_magnetic_reconn, Zhang2017_magnetic_reconnection_kink_instability}; and either of the two, e.g., \citealt{Potter2018_flares}). We do not closely compare the predictions of these models to our BBHOP flares, because such model comparisons require careful flare selection (e.g., \citealt{Jormanainen2018_MscThesis}), which is beyond the scope of this work involving tens of thousands of flares.

\subsubsection{BBHOP flare characteristics against synchrotron peak frequency} \label{sec_results_flares_v_nu}
In Fig. \ref{fig_LogLE_v_SourceAVG_flare_TOP}, we show the duty cycle, source-averaged average rise and fall times (i.e., half of flare duration on average), source-averaged global amplitude, and source-averaged average local amplitudes of the prominent BBHOP flares against synchrotron peak frequency. In Appendix \ref{appendix_distr_of_ALL_flares} we show these plots using all BBHOP flares (see Fig. \ref{fig_LogLE_v_SourceAVG_flare_ALL}). The main trends and results are identical between Figs. \ref{fig_LogLE_v_SourceAVG_flare_TOP} and \ref{fig_LogLE_v_SourceAVG_flare_ALL}.

Regarding BBHOP duty cycles, as seen in Fig. \ref{fig_LogLE_v_SourceAVG_flare_TOP} (i and ii), while a peak at $\nu_\mathrm{sy} \sim 10^{13}$~Hz may be present for duty cycle against $\nu_\mathrm{sy}$, this cannot be confirmed. This is because the duty cycle is mostly flat, albeit very weakly dropping, as $\nu_\mathrm{sy}$ increases (for BLLs $\tau=-0.07$ at $>$3$\sigma$). This almost flat behavior is similar to that of the BB95 fraction against $\nu_\mathrm{sy}$ in Fig. \ref{fig_critically_bright_BB_frac_TOP} (see Sect. \ref{sec_results_critically_bright_BB}). Interestingly, the duty cycle of all BBHOP flares (Fig. \ref{fig_LogLE_v_SourceAVG_flare_ALL} panels i and ii) shows a clearer drop with increasing $\nu_\mathrm{sy}$ (for BLLs $\tau=-0.18$ at $>$3$\sigma$). Such a decreasing trend would generally agree with \cite{Paliya2017_duty_cycle_vs_nu}, who monitored the intranight optical variability of 17 blazars and found that LSPs are more variable and show higher duty cycles than ISPs and HSPs. Nonetheless, we cannot confidently claim a clear decreasing trend in duty cycle against $\nu_\mathrm{sy}$.

From Fig. \ref{fig_LogLE_v_SourceAVG_flare_TOP} (iii and iv), we see that the average rise and fall times for prominent flares is smallest at $\nu_\mathrm{sy} \sim 10^{13}$~Hz with a clear increasing trend toward higher $\nu_\mathrm{sy}$ (for BLLs $\tau=+0.17$ at $>$3$\sigma$). This agrees with the argument presented in Sect. \ref{sec_results_variability_vs_nu}, since probing lower energies than the synchrotron peak frequency means observing particles which have longer cooling times. Additionally, the flaring timescales for FSRQs and BLLs are on average comparable (regardless of $\nu_\mathrm{sy}$), which is similar to their behavior in the radio band (e.g., \citealt{Hovatta2007_radio_longerm_var_analysis}).

Likewise, from Fig. \ref{fig_LogLE_v_SourceAVG_flare_TOP} (v--viii), we see that the global and local amplitudes of the prominent BBHOP flares peak at $\nu_\mathrm{sy} \sim 10^{13}$~Hz and exhibit a clear drop toward higher $\nu_\mathrm{sy}$, as evident by the $>$3$\sigma$ mildly negative $\tau$ correlations of $-0.19$ and $-0.17$ for the BLL global and local amplitudes, respectively, against $\nu_\mathrm{sy}$. A similar behavior with respect to fractional variability was seen in Sect. \ref{sec_results_variability_vs_nu}, where the highest variability occurs at the synchrotron peak frequency. This is also inline with the reasoning presented in Sect. \ref{sec_results_variability_vs_nu}, since emissions corresponding to the synchrotron peak frequency come from particles with the highest energies and thus exhibit the largest flares. Notably, even in terms of flare amplitudes, FSRQs and BLLs at the same $\nu_\mathrm{sy}$ range behave similarly.

The similarities between the flux density, variability, and flaring behaviors of FSRQs and BLLs at the same $\nu_\mathrm{sy}$ range suggest that the processes which dominate their flaring behavior are likely similar. Crucially, this suggests that differences between the variability and flaring behaviors of FSRQs and BLLs in studies where their entire populations are considered (i.e., without limiting $\nu_\mathrm{sy}$; e.g., \citealt{Bauer2009_PTF, Zhang2024_opt_var_not_z_dependent}) generally arise due to the distinct $\nu_\mathrm{sy}$ ranges of FSRQs and BLLs.

\subsubsection{BBHOP flare characteristics against radio variability Doppler factor} \label{sec_results_flares_v_DopF}
In Fig. \ref{fig_DopF_v_SourceAVG_flare_TOP}, we show the duty cycle, source-averaged average rise and fall times, source-averaged global amplitude, and source-averaged average local amplitudes of prominent BBHOP flares against radio variability Doppler factor. These plots for all BBHOP flares are given in Appendix \ref{appendix_distr_of_ALL_flares} Fig. \ref{fig_DopF_v_SourceAVG_flare_ALL}. The main trends and results are identical between Figs. \ref{fig_DopF_v_SourceAVG_flare_TOP} and \ref{fig_DopF_v_SourceAVG_flare_ALL}.

In Fig. \ref{fig_critically_bright_BB_frac_TOP} (i and ii), we see that the duty cycle of the prominent flares remains constant with respect to radio variability Doppler factor (similar to the behavior of BB95 fraction against radio variability Doppler factor; see Fig. \ref{fig_critically_bright_BB_frac_TOP}). This behavior is expected because the frequency of events that lead to flares in blazar jets is not expected to be modified by Doppler factor. Nevertheless, a greater average Doppler factor should imply shortened rise and fall times because of shortened photon arrival timescales in the observer frame (i.e., $\Delta t_\mathrm{obs} \propto 1/\delta$, where $\Delta t_\mathrm{obs}$ and $\delta$ are the observed flaring times and the Doppler factor, respectively; see Eqn. \ref{eqn_flare_time}). Likewise, a greater average Doppler factor should result in larger flare amplitudes because of enhanced flux density in the observer frame (i.e., $\Delta S_\mathrm{obs} \propto \delta^{3-\alpha}$, where $S_\mathrm{obs}$, $\delta$, and $\alpha$ are the observed flux densities at a given frequency, Doppler factor, and spectral index as defined by $S \propto \nu^{\alpha}$, respectively; see Eqn. \ref{eqn_flare_amplitude}).

We see a hint of these trends in Fig. \ref{fig_DopF_v_SourceAVG_flare_TOP} (iii--viii). For FSRQs, which have three times more radio variability Doppler factor estimates than BLLs, we find that the average flaring times show a very weak negative correlation with radio variability Doppler factor ($\tau=-0.09$ at 2.7$\sigma$), and the global and local amplitudes weak positive correlations ($\tau \approx +010$ at $>$3$\sigma$). In Appendix \ref{appendix_z_corr_Doppler_factor} we correct the flare parameters for the effect of redshift and, by extension, luminosity distance. After this correction, the expected flare parameter versus Doppler factor trends become stronger and significant ($>$3$\sigma$), as seen in Fig. \ref{fig_DopF_v_SourceAVG_flare_FlrCorr_res_TOP}. The $>$3$\sigma$ negative correlation between redshift-corrected flare times and radio variability Doppler factor is milder for FSRQs with $\tau=-0.18$ than BLLs with $\tau=-0.31$. Likewise, the $>$3$\sigma$ positive correlation between redshift-corrected flare amplitudes and radio variability Doppler factor is milder for FSRQs with $\tau=+0.19$ than BLLs with $\tau\approx+0.40$. Nevertheless, we caution that these trends may be partially due to host-galaxy contamination for low redshift sources which affects BLLs more strongly (see Appendix \ref{appendix_z_corr_Doppler_factor} for more details). It is noteworthy that such flare time and amplitude trends against variability Doppler factor were also seen by \cite{Hovatta2007_radio_longerm_var_analysis} who analyzed radio flares in 53 blazars.

\section{Summary and conclusion} \label{sec_conclusion}
In this paper we have constructed a catalog of 7918 blazars and blazar candidates, which is the largest of its kind to date. This catalog is referred to as CAZ and was constructed by combining several blazar-dominated AGN catalogs and samples (RFC, 4LAC, 3HSP, CGRaBS, and 5BZC; Sect. \ref{sec_compiling_caz_sources}). The electronic tables accompanying this paper contain the RA and Dec. of the sources and, when available, their source type, redshift, SED class, synchrotron and HE SED peak frequencies, radio variability Doppler factor, and median X-ray flux density (Sect. \ref{sec_compiling_caz_sources}). Their X-ray light curves and spectra as well as an X-ray-selected, statistically complete subsample of the CAZ catalog are found in our companion paper, Paggi et al. (in prep.).

Additionally, in this paper we extracted optical light curves for sources in the CAZ catalog via three all-sky surveys: CRTS, ATLAS, and ZTF (Sect. \ref{sec_lc_extraction}). We cleaned and merged all the filters (Sect. \ref{sec_clean_outliers} and \ref{sec_merge_all_filters}), obtaining light curves that are as long and as dense as possible for as many of the 7918 sources as possible. The merged light curves as well as their corresponding BB data and BBHOP flares are given in the electronic tables.

Furthermore, we first quantified the overall variability of the CAZ light curves via $F_\mathrm{var}$ (see Sect. \ref{sec_variability} for the analysis details, and Sect. \ref{sec_results_variability} for the results). Then, we identified their periods of enhanced emission using critically bright BBs (see Sect. \ref{sec_identifying_critically_bright_BBs} for the analysis details, and Sect. \ref{sec_results_critically_bright_BB} for the results). Subsequently, we identified their flaring periods using BBHOP (see Sect. \ref{sec_identifying_flares} for the analysis details, and Sect. \ref{sec_results_flares} for the results). For some analyses, we only focused on the two main subpopulations of blazars: FSRQs (2495 sources) and BLLs (2726 sources).

Despite the notable increase in the number of blazars and their available parameters, all of our results are generally in agreement with prior studies (as discussed in Sect. \ref{sec_results}). Below is a summary of the results:
\begin{itemize}
    \item Confidently variable CAZ light curves spend $\sim$30\% of the time undergoing prominent flares ($S_\mathrm{peak} > S_{75\%}$), with each flare lasting $\sim$30--130~d.
    \item Confidently variable CAZ light curves spend $\sim$5\% of the time with $S > S_{95\%}$ for typical durations of $\sim$2--13~d.
    \item Optical flares generally have a faster rise than decay.
    \item Blazars are brightest and most variable -- exhibiting higher amplitude and faster flares -- when observed at frequencies close to their $\nu_\mathrm{sy}$. Their brightness and variability decrease as lower and higher frequencies than $\nu_\mathrm{sy}$ are probed.
    \item Optical variability and flare characteristics in FSRQs and BLLs are comparable at the same $\nu_\mathrm{sy}$ range, hinting that their flaring processes are similar.
    \item Optical flare times tend to decrease while flare amplitudes increase as the radio variability Doppler factor increases.
\end{itemize}

The CAZ catalog and light curves allow blazar and, by extension, jetted AGN variability studies to be done with unprecedentedly large sample sizes, which can be vital for understanding various physical phenomena involving SMBH systems and their jets. For example, such long-term, high-cadence, and highly numerous light curves can be used to (1) gain insights into the physics of accretion and properties of the central engine (e.g., \citealt{Kelly2009_opt_var_and_SMBH_mass, Caplar2017_PTF_var_and_accretion, Sun2020_opt_var_vs_accretion_properties}); (2) investigate jet processes and infer the physical properties of the emitting regions (e.g., \citealt{Carini2011_opt_var_and_emission_region_size, Jorstad2013_opt_var_and_jet_features, Marscher2014_TEMZ, Jormanainen2023_MagRec_Mrk421, Jormanainen2025_OJ287, Kankkunen2024_MRO_lognterm_var_analysis}); (3) identify binary SMBH systems or dynamically moving jets (e.g., \citealt{Sillanpaa1988_OJ287, Raiteri2017_twisting_jet, Liodakis2021_moving_jets}); (4) boost the statistical significance of population-based correlation studies (e.g., the elusive blazar-neutrino connection, \citealt{liodakis2022_wild_hunt, Kouch2024_CGRaBS_update, Kouch2025_WH2, Kouch2025_companion_CAZ_v_IC}); (5) understand the origin of the HE emission in blazars through multiwavelength cross-correlation analyses (e.g., \citealt{Fuhrmann2014_fermi_radio_cross_correlation, Ramakrishnan2015_radio_gamma_cross_correlation, lindfors2016, liodakis2018, deJaeger2023_ASSASN_time_lags}); (6) identify counterparts in higher energy observations (e.g., \citealt{Abdo2010_Fermi_gamma_var_analysis, Ruan2012_typical_longtem_opt_var_timescale}); (7) classify blazars (e.g., \citealt{Zywucka2018_identify_blz, Agarwal2023_classify_blz_via_ML}); (8) understand the role of AGN in shaping the evolution of their host galaxy (e.g., \citealt{Hickox2014_AGN_var_and_star_formation}); and (9) prepare for next-generation transient data (e.g., \textit{Vera C. Rubin} Observatory's Legacy Survey of Space and Time; \citealt{Ivezic2019_LSST_sci_cases}).

\section*{Data availability}
The CAZ catalog containing 7918 sources along with their compiled physical parameters as well as the corresponding CAZ light curves, BB information, and BBHOP flare information are available via \href{https://cdsarc.cds.unistra.fr/viz-bin/cat/J/A+A/708/A382}{https://cdsarc.cds.unistra.fr/viz-bin/cat/J/A+A/708/A382}. The catalog provides the J2000 RA and Dec. of the CAZ sources along with their RFC, 4LAC, 3HSP, CGRaBS, and 5BZC names as well as their source type, redshift, synchrotron and high-energy peak frequencies, radio variability Doppler factor, median X-ray flux density, fractional variability in the CRTS and ZTF portions of the CAZ light curve, their CAZ variability status, BB95 fraction, and prominent BBHOP duty cycle. The CAZ light curve of each source, if available, is given as MJD, merged flux density, flux density error, filter, shift factor, and merge information (i.e., which filters are merged together). The BB information of each source, if available, is given as MJD of the left and right edges of a BB, average flux density within the BB and its error, number of data points within the BB, and if the BB is considered a BB95. The BBHOP flare information of each source, if available, is given as the MJD of the start, peak, and end of a BBHOP flare along with the respective average flux density and flux density error at each of those MJDs, along with the information if the BBHOP flare is a prominent one.

\begin{acknowledgements}
We thank the anonymous referee for constructive comments.
We also thank Matteo Cerruti for fruitful discussion on flare symmetry.
PK was supported by the Research Council of Finland projects 346071 and 345899.
EL was supported by the Research Council of Finland projects 317636, 320045, and 346071.
TH was supported by the Research Council of Finland projects 317383, 320085, 345899, and 362571 and the European Union ERC-2024-COG - PARTICLES - 101169986.
JJ was supported by the Research Council of Finland projects 320085 and 345899. 
KK acknowledges support from the European Research Council (ERC) under the European Union's Horizon 2020 research and innovation programme (grant agreement No. 101002352).
IL and AP were funded by the European Union ERC-2022-STG - BOOTES - 101076343. Views and opinions expressed are however those of the author(s) only and do not necessarily reflect those of the European Union or the European Research Council Executive Agency. Neither the European Union nor the granting authority can be held responsible for them.
Based on observations obtained with the Samuel Oschin Telescope 48-inch and the 60-inch Telescope at the Palomar Observatory as part of the \textit{Zwicky} Transient Facility project. ZTF is supported by the National Science Foundation under Grant No. AST-2034437 and a collaboration including Caltech, IPAC, the Weizmann Institute for Science, the Oskar Klein Center at Stockholm University, the University of Maryland, Deutsches Elektronen-Synchrotron and Humboldt University, the TANGO Consortium of Taiwan, the University of Wisconsin at Milwaukee, Trinity College Dublin, Lawrence Livermore National Laboratories, and IN2P3, France. Operations are conducted by COO, IPAC, and UW. The ZTF forced-photometry service was funded under the Heising-Simons Foundation grant \#12540303 (PI: M.J.Graham).
This work has made use of data from the Asteroid Terrestrial-impact Last Alert System (ATLAS) project. The Asteroid Terrestrial-impact Last Alert System (ATLAS) project is primarily funded to search for near earth asteroids through NASA grants NN12AR55G, 80NSSC18K0284, and 80NSSC18K1575; byproducts of the NEO search include images and catalogs from the survey area. This work was partially funded by Kepler/K2 grant J1944/80NSSC19K0112 and HST GO-15889, and STFC grants ST/T000198/1 and ST/S006109/1. The ATLAS science products have been made possible through the contributions of the University of Hawaii Institute for Astronomy, the Queen's University Belfast, the Space Telescope Science Institute, the South African Astronomical Observatory, and The Millennium Institute of Astrophysics (MAS), Chile.
This work has made use of data from the Joan Or\'o Telescope (TJO) of the Montsec Observatory (OdM), which is owned by the Catalan Government and operated by the Institute for Space Studies of Catalonia (IEEC).
This work makes use of Matplotlib (\citealt{Hunter2007_Matplotlib}), NumPy (\citealt{Harris2020_NumPy}), SciPy (\citealt{Virtanen2020_SciPy}), and Astropy (\citealt{Astropy2022_v5}).
\end{acknowledgements}

\bibliographystyle{aa} 
\bibliography{ref.bib} 

\begin{appendix}

\onecolumn
\section{Variability measures other than $F_\mathrm{var}$} \label{appendix_other_var}

\begin{figure*} [h!]
    \centering
    \includegraphics[width=17cm]{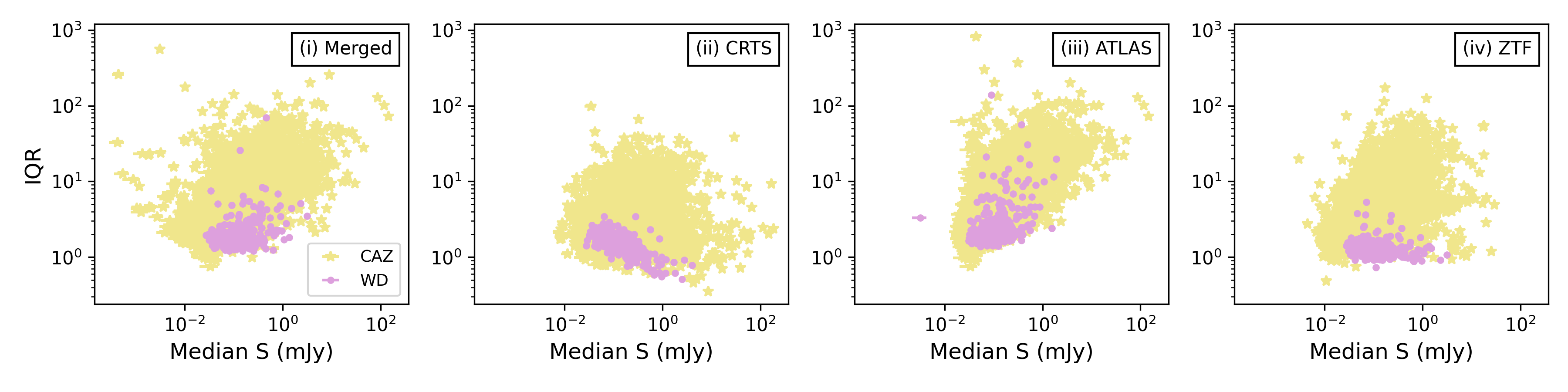}
    \caption{IQR versus median flux density plots. Plots (i--iv) show the scatter for merged light curves, CRTS-, ATLAS-, and ZTF-only, respectively.}
    \label{fig_IQR_v_flux}
\end{figure*}

\begin{figure*} [h!]
    \centering
    \includegraphics[width=17cm]{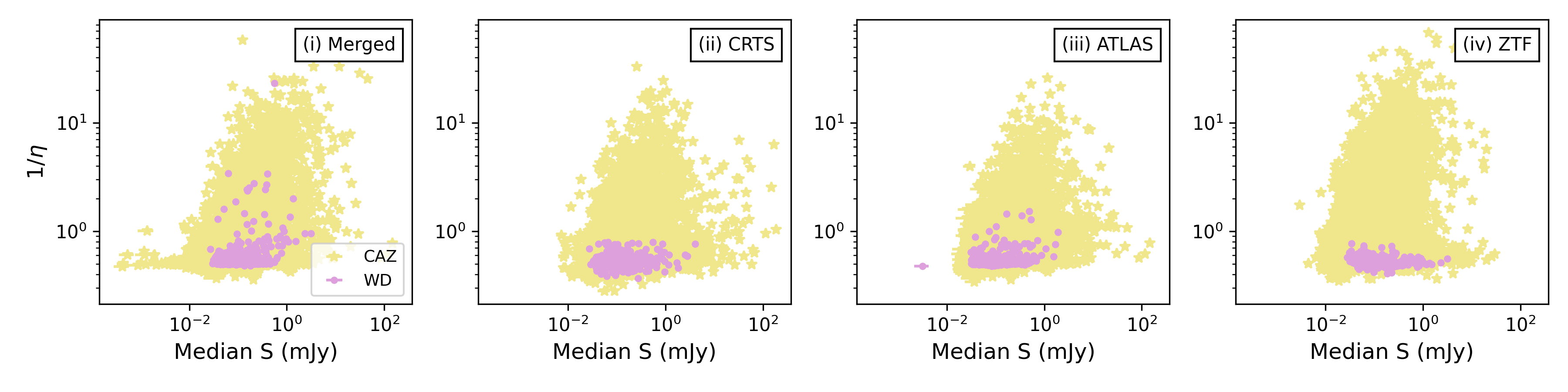}
    \caption{$1/\eta$ versus median flux density plots. Plots (i--iv) show the scatter for merged light curves, CRTS-, ATLAS-, and ZTF-only, respectively.}
    \label{fig_OoE_v_flux}
\end{figure*}

\begin{figure*} [h!]
    \centering
    \includegraphics[width=17cm]{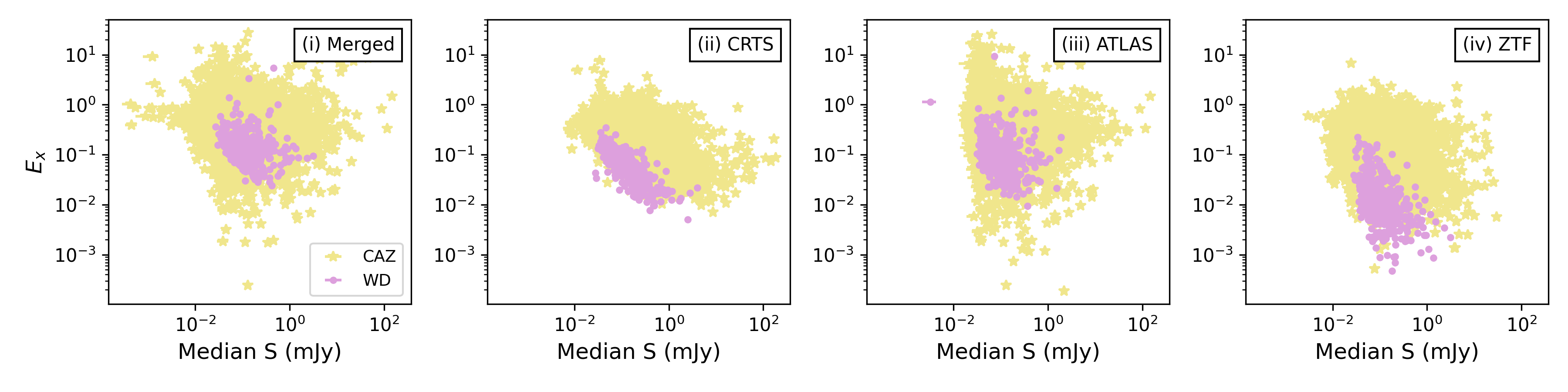}
    \caption{$E_x$ versus median flux density plots. Plots (i--iv) show the scatter for merged light curves, CRTS-, ATLAS-, and ZTF-only, respectively.}
    \label{fig_Exc_v_flux}
\end{figure*}

Here we describe three metrics other than $F_\mathrm{var}$, namely, IQR, $1/\eta$, and $E_x$, which we consider in this study for quantifying the overall variability of the CAZ light curves. In Sect. \ref{sec_variability} we describe why we opt to focus on $F_\mathrm{var}$ as the main metric. Below we show the variability versus flux density plots for IQR, $1/\eta$, and $E_x$ in Figs. \ref{fig_IQR_v_flux}, \ref{fig_OoE_v_flux}, and \ref{fig_Exc_v_flux}, respectively.

IQR is a scatter-based variability metric, equal to the difference between the 75$^\mathrm{th}$ and 25$^\mathrm{th}$ percentiles of the flux density distribution. We calculate it via $\mathrm{IQR} = (S_\mathrm{75\%}-S_\mathrm{25\%})/\mathrm{MED}(\sigma_i)$, where $\mathrm{MED}(\sigma_i)$ is the median flux density error and acts as a normalization factor. IQR is highly effective in distinguishing between variable and non-variable stars (\citealt{sokolovsky2016}).

The \cite{vonNeumann1941_eta} ratio, $\eta$, measures the mean square successive difference divided by the variance. Since variance increases with variability, a larger inverse ($1/\eta$) indicates a larger degree of variability (e.g., \citealt{Shin2009_one_over_eta}). This is calculated via $1/\eta = \sum_{i=1}^N (S_i-\overline{S})^2 / \sum_{i=1}^{N-1} (S_{i+1}-S_i)^2$, where $\overline{S}$ is the sample mean. Notably, $1/\eta$ is a correlation-based quantity which measures variability while taking the temporal order into account. 

$E_x$, another correlation-based variability metric, considers the observational gaps of a light curve (\citealt{sokolovsky2016}). We divide the CAZ light curves into observational seasons by scanning for gaps longer than 60~d. Thus, $E_x$ is calculated as:
\begin{equation} \label{eqn_excursions}
    E_x = \frac{2}{N_\mathrm{s}(N_\mathrm{s}-1)} \sum_{i=1}^{N_\mathrm{s}-1} \sum_{j>i}^{N_\mathrm{s}} \frac{\big| \mathrm{MED}(S_{\mathrm{s},i}) - \mathrm{MED}(S_{\mathrm{s},j})  \big|}{\sqrt{\mathrm{MAD}(S_{\mathrm{s},i})^2 + \mathrm{MAD}(S_{\mathrm{s},j})^2}}
\end{equation}
where $N_s$ refers to the number of seasons and MAD is the median absolute deviation, calculated via $\mathrm{MAD}(S_{\mathrm{s},i}) = \mathrm{MED} (| S_{\mathrm{s},i} - \mathrm{MED}(S_{\mathrm{s},i})|)$ where $S_{\mathrm{s},i}$ denotes flux densities within season $i$ (likewise for season $j$).

\twocolumn
\section{Calibrating Bayesian blocks} \label{appendix_BB_calib}
As mentioned in Sect. \ref{sec_identifying_flares}, CAZ light curves of 408 WDs are used to calibrate \texttt{ncp\_prior} of the BB algorithm in order to minimize the effect of noise. This is done by incrementally increasing \texttt{ncp\_prior}$_\mathrm{ideal}$~=~$1.32 + 0.577 \log_{10}(N)$ until the resulting BB light curve (with a total of $N$ flux densities) is mostly consistent with a constant flux density corresponding to the median of the merged light curve. We note that CRTS and other (i.e., non-CRTS) filters are treated separately in this procedure.

We define an extreme WD BB fraction as the length of all BB whose flux density does not reach the median of the merged light curve within $2\sigma$, divided by the total length of all BB. In case of ideal Gaussian errors, we would expect this extreme fraction to be $\sim$5\% because 2$\sigma$ Gaussian error bars should signify a $\sim$95\%-likelihood range for the constant WD flux densities. However, in the case of WD CAZ light curves, \texttt{ncp\_prior}$_\mathrm{ideal}$ generally results in extreme fractions greater than 5\%. This demonstrates the noisy nature of the CAZ light curves, as mentioned multiple times throughout this paper.

\begin{figure} [h!]
    \centering
    \includegraphics[width=0.49\textwidth]{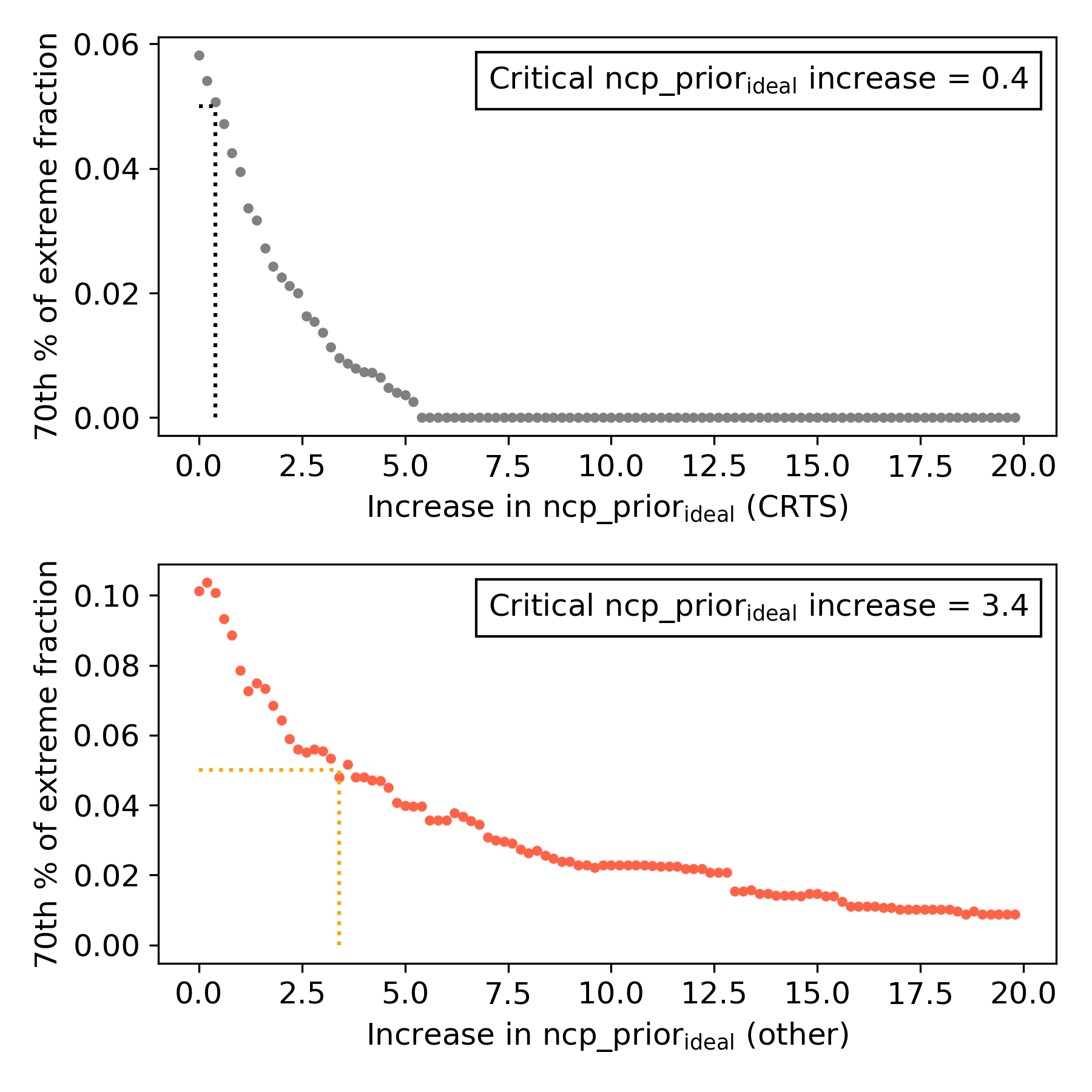}
    \caption{Value containing $\sim$1$\sigma$ (70\%) of the extreme WD BB fractions at each incremental increase of \texttt{ncp\_prior}$_\mathrm{ideal}$. \textit{Top}: CRTS WD light curves; \textit{bottom}: other (i.e., non-CRTS) WD CAZ light curves. The dotted lines show the critical value of the increase in \texttt{ncp\_prior}$_\mathrm{ideal}$ at the 0.05 level. It is 0.4 and 3.4 for CRTS and other filters, respectively.}
    \label{fig_calib_BB}
\end{figure}

To quantify the optimal value of \texttt{ncp\_prior}, we incrementally add 0.2 to its ideal value (only dependent on $N$) for each 408 WD light curves. We do this 100 times, covering a range of \texttt{ncp\_prior}~=~\texttt{ncp\_prior}$_\mathrm{ideal}$ to \texttt{ncp\_prior}~=~\texttt{ncp\_prior}$_\mathrm{ideal}+20$. For each step, we calculate the value within which $\sim$1$\sigma$ (70\%) of the 408 extreme WD BB fractions are contained. This is plotted versus the incremental increase in \texttt{ncp\_prior}$_\mathrm{ideal}$ for both CRTS and other filters separately in Fig. \ref{fig_calib_BB}. The calibrated increase in \texttt{ncp\_prior}$_\mathrm{ideal}$ is defined as the value that results in $\sim$1$\sigma$ of the extreme fractions to be 0.05. This is 0.4 for CRTS, and 3.4 for other (non-CRTS) filters.

This result suggests that CRTS light curves are less noisy than the other filter light curves, which we confirm by visual inspection. ATLAS and ZTF light curves are especially prone to extreme outliers, noise-like behavior, and underestimated errors. This is why we select the global marker only at the $\sim$1$\sigma$ level, instead of a stricter one. For example, even as \texttt{ncp\_prior}~$\rightarrow \infty$, the extreme WD BB fraction of the non-CRTS light curves at the $2\sigma$ level never reaches 0.05 (see the bottom plot in Fig. \ref{fig_calib_BB}).

\onecolumn
\section{Characteristics of all BBHOP flares against physical parameters} \label{appendix_distr_of_ALL_flares}
As discussed previously (see Sect. \ref{sec_identifying_flares} and \ref{sec_results_flares}), there are many noisy BBHOP flares. Thus, when analyzing their characteristics against synchrotron peak frequency and radio variability Doppler factor in Sect. \ref{sec_results_flares}, we only focused on a subset of prominent BBHOP flares (see Figs. \ref{fig_LogLE_v_SourceAVG_flare_TOP} and \ref{fig_DopF_v_SourceAVG_flare_TOP}). In this appendix we provide the same distributions for all BBHOP flares (not just prominent ones).

\begin{figure*}[h!]
    \centering
    \includegraphics[width=17cm]{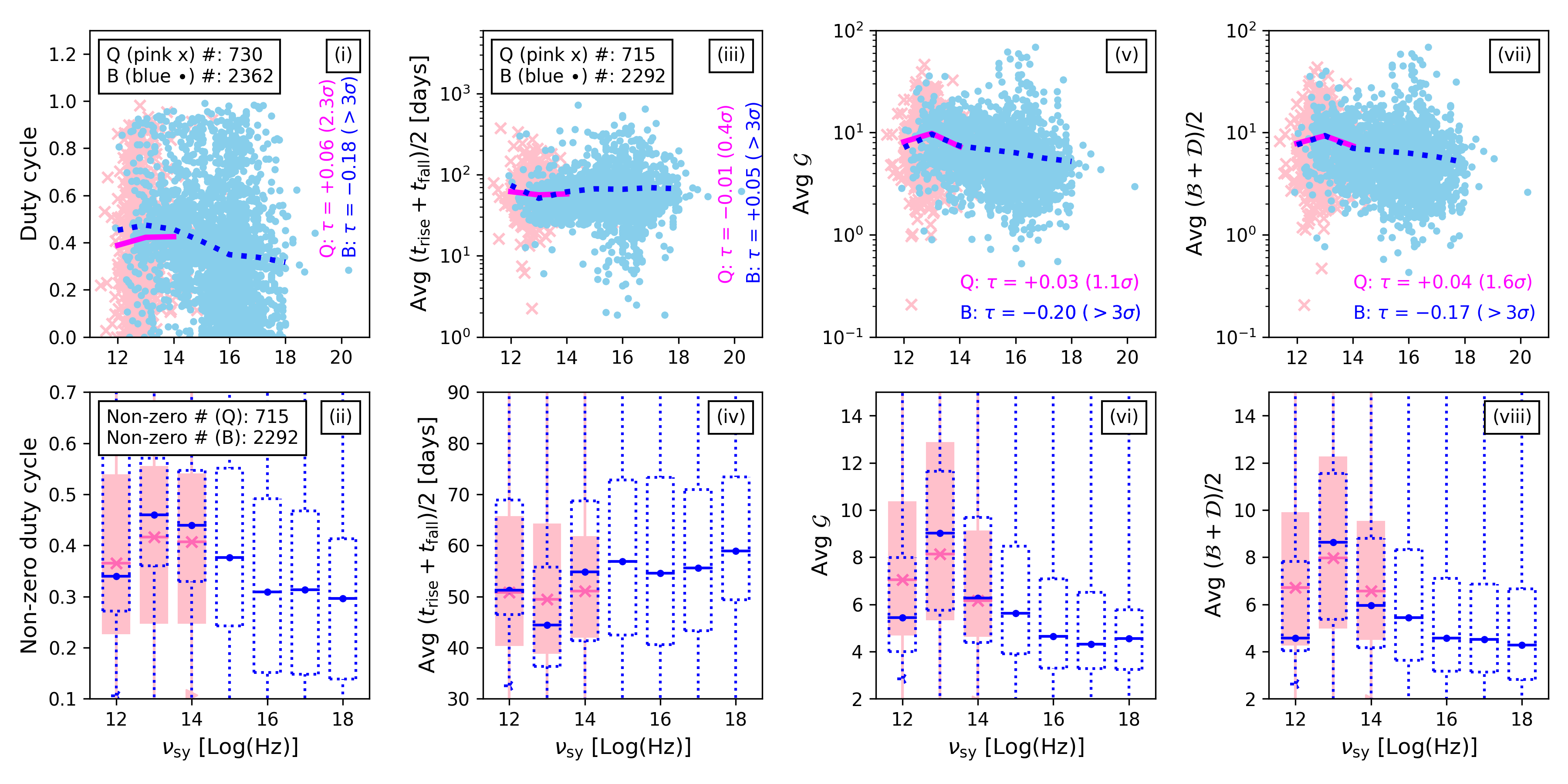}
    \caption{Distribution of source duty cycle (i and ii) as well as source-averaged average rise and fall times (iii and iv), global amplitude (v and vi), and average local brightening and decaying amplitudes (vii and viii) of all BBHOP flares against synchrotron peak frequency. Pink crosses and solid markings refer to FSRQs (Q); blue dots and dotted markings to BLLs (B). Kendall's $\tau$ correlation coefficient and significance are reported for FSRQs and BLLs. In plot (i) we show zero duty cycle data, whereas in the other plots we do not. The running averages in the top row (shown as solid lines for FSRQs and dotted lines for BLLs) and the box plots in bottom row are both only calculated for at least ten nonzero values within each bin. In the top panels we show all data points, while in the bottom panels we zoom in on the boxes.}
    \label{fig_LogLE_v_SourceAVG_flare_ALL}
\end{figure*}

\begin{figure*}[h!]
    \centering
    \includegraphics[width=17cm]{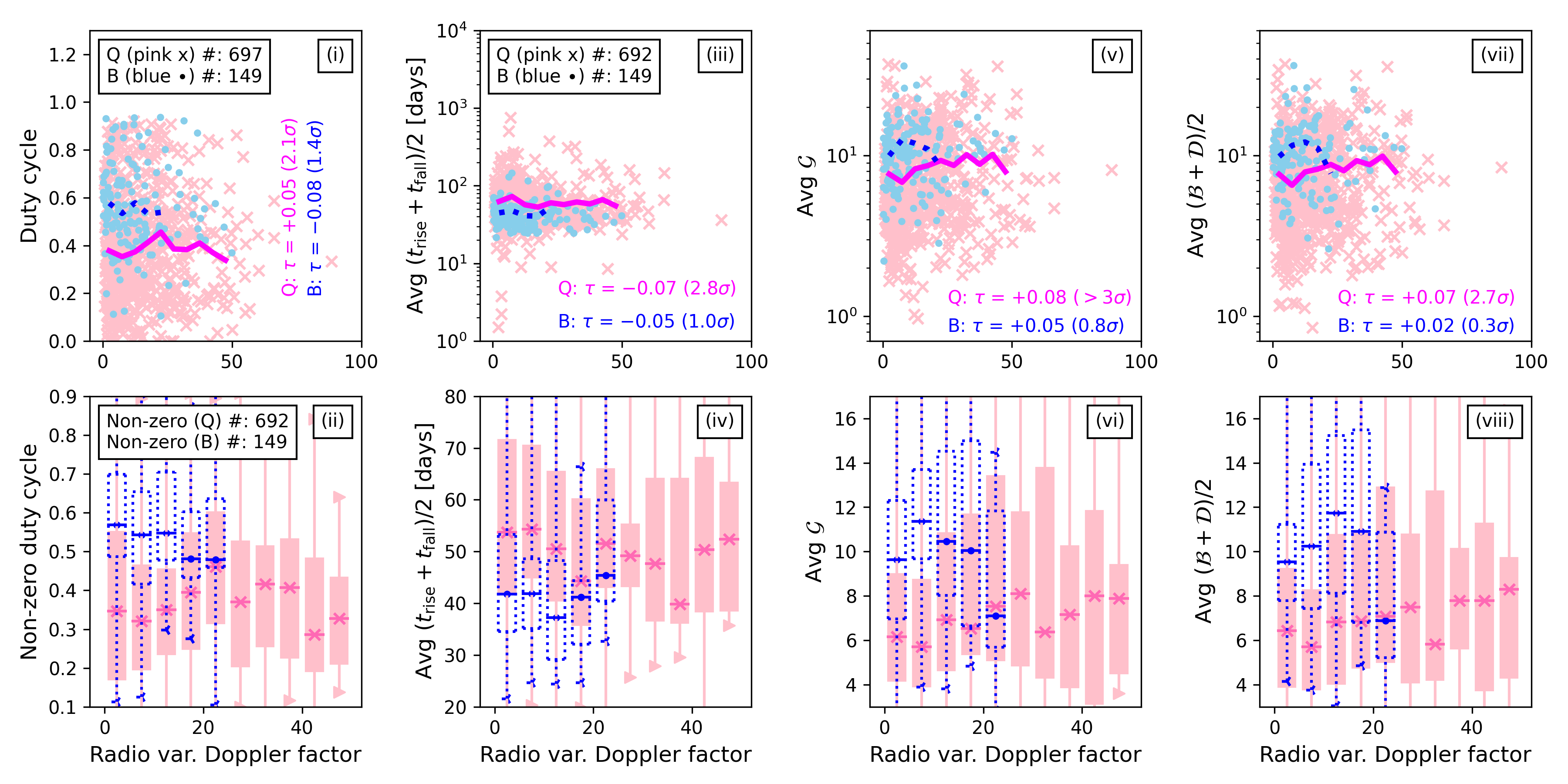}
    \caption{Similar as Fig. \ref{fig_LogLE_v_SourceAVG_flare_ALL} but for radio variability Doppler factor.}
    \label{fig_DopF_v_SourceAVG_flare_ALL}
\end{figure*}

\onecolumn
\section{Redshift-corrected flare parameters against radio variability Doppler factor} \label{appendix_z_corr_Doppler_factor}
In Sect. \ref{sec_results_flares_v_DopF} Fig. \ref{fig_DopF_v_SourceAVG_flare_TOP} we investigate the behavior of duty cycle as well as source-averaged average rise and fall times, and local and global amplitudes of prominent BBHOP flares against radio variability Doppler factor. While rise and fall times seem to decrease and amplitudes increase with increasing Doppler factor, here we investigate how correcting the flare parameters for redshift and luminosity distance changes the trends. We note that duty cycle (i.e., frequency of flaring) is not affected by redshift and, as such, is omitted from the investigation of this appendix.

As given in \cite{Kembhavi1999_AGN_book}, the intrinsic flare times are calculated via:
\begin{equation} \label{eqn_flare_time}
    \Delta t_\mathrm{int} = \Delta t_\mathrm{obs} \cdot \left(\frac{\delta}{1+z} \right)
\end{equation}
where $\Delta t_\mathrm{int}$, $\Delta t_\mathrm{obs}$, $\delta$, and $z$ are the source-frame flaring rise or fall time, the observed flaring rise or fall time, the Doppler factor, and the redshift, respectively. Meanwhile, the intrinsic flaring amplitude (in terms of luminosity) are calculated via:
\begin{equation} \label{eqn_flare_amplitude}
    \Delta L_\mathrm{int} = \Delta S_\mathrm{obs} \cdot \left( \frac{4 \pi {d_\mathrm{L}}^2}{1+z} \right) \cdot \left( \frac{1+z}{\delta} \right)^{3- \alpha}
\end{equation}
where $\Delta L_\mathrm{int}$, $\Delta S_\mathrm{obs}$, $d_\mathrm{L}$, and $\alpha$ are the source-frame flaring amplitude in terms of luminosity, the observed flaring amplitude in terms of flux density, luminosity distance, and spectral index as defined by $S \propto \nu^{\alpha}$, respectively. Spectral index in the optical band is class dependent. Here we use LSPs have $\alpha = - 1.1$, ISPs $\alpha = -1.3$, and HSPs $\alpha = -1.5$ (\citealt{Fiorucci2004_spectral_index, Hovatta2014_opt_vs_gamma_variability}). 

\begin{figure*}[h!]
    \centering
    \includegraphics[width=16.5cm]{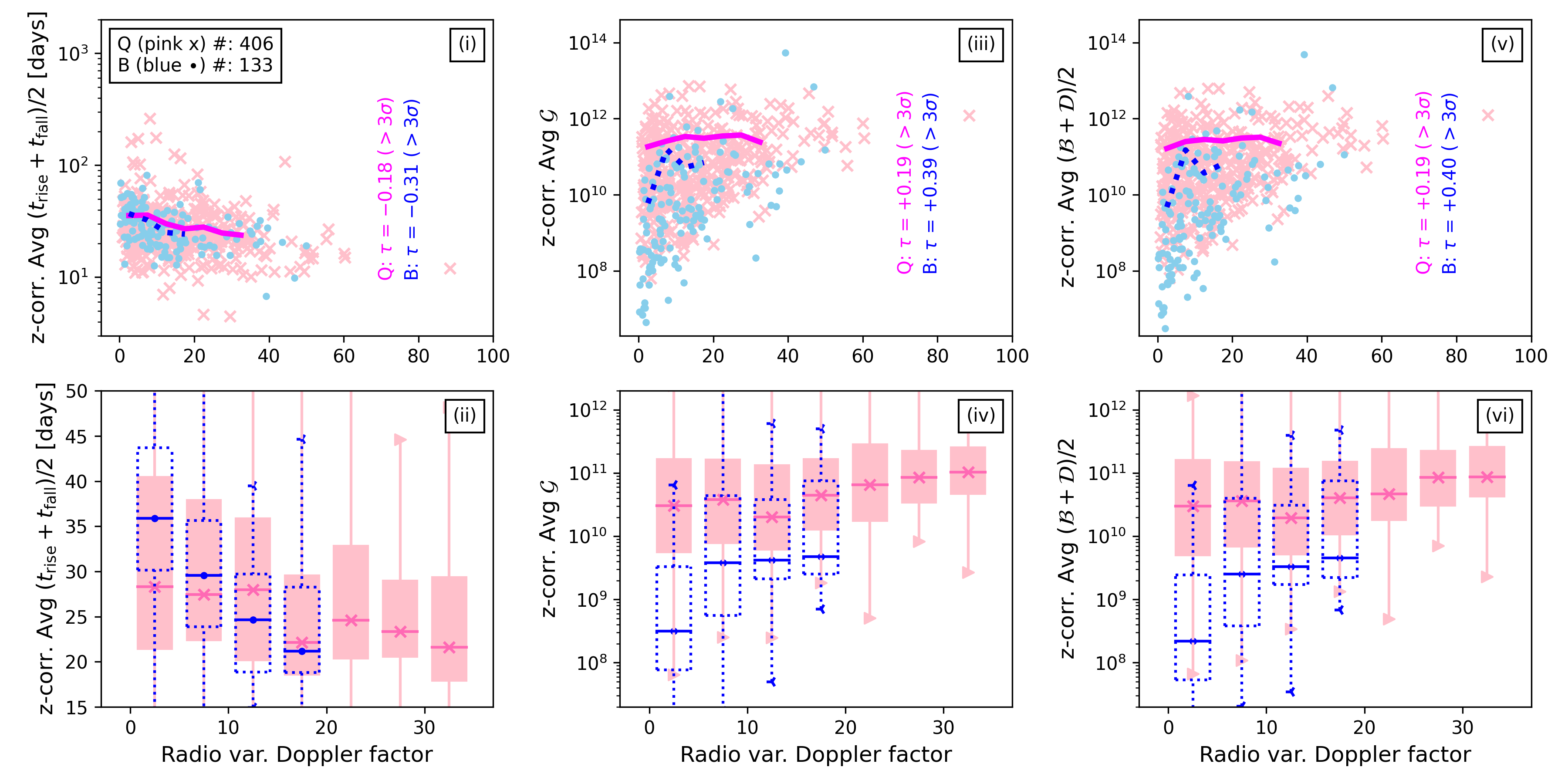}
    \caption{Distribution of redshift-corrected, source-averaged average flaring times (i and ii), global amplitudes (iii and iv), and average local amplitudes (v and vi) of prominent BBHOP flares against radio variability Doppler factor. For more plot details, see Fig. \ref{fig_DopF_v_SourceAVG_flare_TOP} where we show the same trends prior to redshift-correction.}
    \label{fig_DopF_v_SourceAVG_flare_FlrCorr_res_TOP}
\end{figure*}

Since we correct the flare parameters with respect to redshift and luminosity distance, and then compare them to radio variability Doppler factor estimates, in Eqns. \ref{eqn_flare_time} and \ref{eqn_flare_amplitude}, we only focus on $z$ and $d_\mathrm{L}$ while ignoring $\delta$. We divide the source-averaged average rise and fall times of the prominent BBHOP flares by $1+z$, and multiply their source averaged global and local amplitudes (which are proportional to the global and local $\Delta S_\mathrm{obs}$, respectively) by $4 \pi {d_\mathrm{L}}^2/(1+z)^{2-\alpha}$. These redshift-corrected (``z-corr.'') flare parameters against radio variability Doppler factor are plotted in Fig. \ref{fig_DopF_v_SourceAVG_flare_FlrCorr_res_TOP}. We note that to calculate $d_\mathrm{L}$ from $z$, we use the \texttt{cosmology} subpackage of \texttt{astropy} by assuming the cosmological constants of \cite{Plack2014_astropy_cosmology}.

As evident from Fig. \ref{fig_DopF_v_SourceAVG_flare_FlrCorr_res_TOP}, the redshift-corrected flaring times become shorter and amplitudes larger as the radio variability Doppler factor increases. The $>$3$\sigma$ negative correlation of redshift-corrected flare times with respect to radio variability Doppler factor is $\tau=-0.18$ for FSRQs and $\tau=-0.31$ for BLLs. Meanwhile, the $>$3$\sigma$ positive correlation of redshift-corrected flare amplitudes with respect to radio variability Doppler factor is $\tau=+0.19$ for FSRQs and $\tau \approx +0.40$ for BLLs.

Notably, in both cases, the trends are more envelope-like than linear. At lower radio variability Doppler factors, a wider range of flare parameters are possible: faster and larger-amplitude flares as well as slower and lower-amplitude ones. However, at higher radio variability Doppler factors, there are fewer slow and low-amplitude flares. While this may be physical, it should be noted that it is, at least partially, because of host-galaxy contamination. Generally, blazars with lower Doppler factors are dimmer with respect to their host-galaxy and thus are more strongly affected by host-galaxy contamination, which diminishes their variability and our ability to identify prominent flares. This is likely why the trends are stronger in the case of BLLs than FSRQs, because BLLs are more strongly affected by host-galaxy contamination than FSRQs.

\end{appendix}

\end{document}